%% file: R-est-v18.tex
\documentclass[a4paper,11pt]{article}
\usepackage[a4paper, text={16cm,25cm}]{geometry}
\usepackage{parskip}
\usepackage{amsmath}
\usepackage{amsthm}
\usepackage{natbib}
\usepackage{graphicx}
\usepackage{hyperref}
\usepackage{appendix}
\usepackage{bbm}
\newcommand{\blind}{1}
\newcommand{\jrssa}{0}
\usepackage{placeins}

\newcommand{\mb}[1]{\boldsymbol{#1}}
\newcommand{\Var}{\textrm{Var}}
\newcommand{\diag}{\textrm{diag}}
\newcommand{\Cov}{\textrm{Cov}}
\newcommand{\N}{\mathcal{N}}
\newtheorem{lemma}{Lemma}

\title{Simultaneous estimation of the effective reproduction number and the
time series of daily infections: Application to Covid-19}
\author{Hans R. K\"unsch and Fabio Sigrist\if1\jrssa{\thanks{Corresponding author. Email: fabio.sigrist@stat.math.ethz.ch. Address: Fabio Sigrist, Seminar for Statistics, ETH Z\"urich, Rämistrasse 101, 8092 Zurich, Switzerland. ORCID identifier: 0000-0002-3994-2244}}\fi\\Seminar f\"ur Statistik, ETH
  Z\"urich }

\begin{document}

\maketitle
\begin{abstract}
The time-varying effective reproduction number is an important parameter for communication and policy decisions during an epidemic. In this paper, we present new statistical methods for estimating the reproduction number based on the popular model of \citet{cori2013new} which defines the effective reproduction number based on self-exciting dynamics of new infections. Such a model is conceptually simple and less susceptible to misspecifications than more complicated multi-compartment models. However, statistical inference is challenging, and the previous literature has either relied on proxy data and/or a two-step approach in which the number of infections is first estimated. In contrast, we present a coherent Bayesian method that approximates the joint posterior of daily new infections and reproduction numbers using a novel Markov chain Monte Carlo (MCMC) algorithm. Comparing our method to the state-of-the-art three-step estimation procedure of \citet{huisman2022estimation}, both using daily confirmed cases from Switzerland in the Covid-19 epidemic and simulated data, we find that our method is more accurate in terms of point estimates and uncertainty quantification, especially near the beginning and end of an observation period. 
\end{abstract}

\if1\jrssa{{\it Keywords:} Bayesian inference, infectious disease modeling, Markov chain Monte Carlo, non-stationary self-exciting Poisson process  }\fi

\input{main.tex}

\section*{Acknowledgements}

We thank Markus Petermann and Daniel Wyler for comments and encouragement at difficult stages of the project. We also thank two anonymous reviewers and the associate editor for their constructive comments, which helped to improve this article.

\bibliographystyle{abbrvnat}
\bibliography{bib_r_est.bib}

\clearpage

\include{appendix.tex}

\end{document}

%% file: main.tex
\section{Introduction}
\label{intro}

The effective reproductive number is the average number of secondary infected cases per infected case in a population, taking into account variations in time of the percentage of susceptible persons in the population, the infectivity of the disease, and the number of contacts between infectious and susceptible members of the population. It is used in many countries as an indicator of the growth of an epidemic and as a basis for control policies.

The daily effective reproduction numbers $R_{e,t}$ cannot be measured directly, but have to be estimated based on a model for the dynamics of an epidemic and on available data such as daily numbers of confirmed cases, hospitalizations or deaths. Various methods have been developed to estimate $R_{e,t}$ which can be broadly classified into two categories: Multi-compartment model-based approaches \citep[e.g.,][]{delamater2019complexity,kucharski2020early,zhou2020preliminary}, and methods which directly estimate the number of secondary infections per case using infection incidence time series \citep[e.g.,][]{wallinga2007generation, cori2013new}. We focus on the latter type of methods which obtain the effective reproduction numbers using a model for the daily new infections. Compared with other methods, ``they are  based on fewer and simpler assumptions, less susceptible to model misspecifications, and well-suited for real-time epidemic monitoring'' \citep{gostic2020practical}.

Specifically, we consider the model of \citet{cori2013new}, which is widely used \citep{brockhaus2023different} and models the unobserved number of daily infections as a self-exciting Poisson process with a rate proportional to $R_{e,t}$. If the number of infections $I_t$ were available, estimation of $R_{e,t}$ by maximum likelihood or by a Bayesian method would be straightforward; see \citet{cori2013new}. However, the daily new infections are typically unobserved, making estimation difficult. Existing methods either replace $I_t$ by proxies like the onset of symptoms, by estimates $\hat I_t$ obtained from the daily detected cases $D_t$, or by a combination of both. The use of symptom-onset data is problematic for modeling, e.g., Covid-19 since persons can be infectious even without symptoms \citep{rothe2020transmission}, though. If the delays between infection and detection are i.i.d. random, then the conditional expectation $E[D_t \mid I_t, I_{t-1},...]$ is obtained as a convolution of $I_t$ with the detection distribution. Therefore, estimates $\hat{I}_t$ can be obtained by deconvolution methods, see \citet{goldstein2009reconstructing}.

However, estimating the effective reproduction numbers in a second step from such proxies or from deconvolution estimates $\hat{I}_t$ instead of the true number of daily infections $I_t$ ignores an essential component of uncertainty. \citet{goldstein2009reconstructing} therefore suggested bootstrap methods to take this uncertainty into account. Similarly, \citet{huisman2022estimation} proposed a residual block bootstrap procedure which can handle dependence and non-stationarity of the detections $D_t$. This allows to compute not only point, but also interval estimates. Their method was then used by the Swiss federal office of public health during the Covid-19 pandemic.

Besides the difficulty of obtaining accurate uncertainty estimates, such a two-step approach entails multiple problems. First, the deconvolution method in \citet{goldstein2009reconstructing}, called Richardson-Lucy deconvolution, is an instance of the EM-algorithm. The function which is maximized by the EM-algorithm has many maxima and the result depends on the starting value, see the Supplementary Material \ref{app-em} for an example. Moreover, the EM-algorithm is based on assumptions that are not compatible with the self-exciting Poisson process model for the daily infections of \citet{cori2013new}, see the comment at the beginning of Section \ref{methods}. 

In this article, we introduce methodology that allows to jointly estimate the number of infections $I_t$ and the reproduction numbers $R_{e,t}$ and to obtain coherent uncertainty estimates for the model of \citet{cori2013new} without relying either on proxy data or on a two-step approach. It consists of a Markov chain Monte Carlo (MCMC) algorithm which samples from the joint posterior of $I_t$ and $R_{e,t}$ given $D_t$. We also address sequential estimation and problems due to weekday effects in $D_t$, and we show how our methods can be adapted to extensions of the model by \citet{cori2013new} including over-dispersion in the self-exciting dynamics. Finally, we provide evidence that it provides more accurate results in simulation examples. We believe that our method will be a useful tool for future epidemics and other diseases. 

The outbreak of the Covid-19 pandemic has led to many publications which propose other methods to estimate the effective reproduction numbers. An alternative way to bypass the estimation difficulty inherent in the model of \citet{cori2013new} is to replace their model with another one for which statistical inference is easier \citep[e.g.,][]{abbott2020estimating, flaxman2020estimating, bhatt2023semi, goldstein2024incorporating}. For instance, \citet{abbott2020estimating} and \citet{flaxman2020estimating} use models that look similar to the model of \citet{cori2013new}. However, their dynamic model is deterministic whereas that of \citet{cori2013new} is stochastic and, consequently, the two modeling approaches are very different. In particular, in Section \ref{sec_notation}, we show that the random fluctuations are much larger in the stochastic model compared to the deterministic case. Furthermore, \citet{goldstein2024incorporating} replace the Poisson distribution in the self-exciting dynamics of \citet{cori2013new} by a continuous gamma distribution for modeling the discrete variable $I_t$. Inference is then much simpler as Hamiltonian Monte Carlo methods can be used for continuous distributions. \citet{birrell2020efficient} and \citet{storvik2023sequential} are examples of methods based on compartmental models. They split the population into compartments like susceptible, exposed, infectious and removed (SEIR) and consider random transitions between these compartments. Often, the compartments are split further according to age or smaller geographical units.

Our paper is structured as follows. In Section \ref{sec_notation}, we precisely define the model we use. Section \ref{methods} explains our MCMC algorithm which iterates between updating $R_{e,t}$ given $I_t$ and updating $I_t$ given $R_{e,t}$ and $D_t$. The latter is the bottleneck, and it is sensitive to the details of the implementation. In Section \ref{extension}, we show how our methods can be extended to account for over-dispersion in the self-exciting infection dynamics and to model the detection parameters stochastically using a prior. Section \ref{examples} applies our method to daily positive test results in Switzerland from August 31, 2020, to July 4, 2021, and assesses it in a small simulation experiment. We compare our results with those obtained using the methods of \citet{huisman2022estimation} and study the sensitivity to the treatment of weekday and holiday effects and to the chosen prior for $R_{e,t}$. Finally, we illustrate the behavior of the sequential implementation for the period from August 31 to November 8, 2020, looking at the effect of a new observation on estimates in the past and on predictions. The paper closes with a short discussion and conclusion.

\section{Notation and model assumptions}\label{sec_notation}

The three most important variables are the daily numbers of newly infected individuals $I_t$, the daily effective reproduction numbers $R_{e,t}$, and the daily numbers of newly detected cases $D_t$. In addition, we also use the numbers $A_{s,t}$ of individuals infected on day $s$ and detected on day $t$ and the numbers $A_{s,+}$ of individuals infected on day $s$ which remain undetected. For $s \leq t$ we use the shorthand notations $s\!:\!t=\{s,s+1,...,t\}$ and $x_{s:t}=(x_s, x_{s+1}, ..., x_t)$.

\subsection{Modeling disease transmission}

For disease transmission, we use a standard self-exciting Poisson model controlled by the effective reproduction number. An individual infected at time $s$ infects others on day $t$ with intensity $R_e(t) w_{t-s}$ where $(w_k)$ is the infectivity profile. For identifiability, it is assumed that the $w_k$ sum to one. Moreover, infected individuals are assumed to be no longer infective after $K_w$ days, i.e., $w_k=0$ for $k> K_w$. Infected individuals transmit the disease independently of each other, which means their rates add. From these assumptions, the conditional distribution of the number of infections on day $t$ given the number of infections on days $s<t$ follows:
\begin{equation}\label{eq:model-I}
  I_t \mid (I_s; s<t) \sim \textrm{Poisson}(\lambda_t), \quad
  \lambda_t = R_{e,t}\sum_{k=1}^{K_w} w_k I_{t-k}
\end{equation}
as in \citet{cori2013new}.

The next lemma gives the connection between a time invariant effective reproduction number and the exponential growth rate of infections. The latter depends also on the infectivity profile and is typically more important for communication and policy decisions.

\begin{lemma} If \eqref{eq:model-I} holds with all $w_k > 0$ and constant effective reproduction number $R_e$, then for any initial condition 
$(E[I_s]) \neq 0; s=1,..., K_w)$ 
  $E[I_t] \rho^{-t}$ converges to a positive constant as $t \rightarrow \infty$ where $\rho$ is the unique solution of the equation
  $$\frac{1}{R_e} = \sum_{k=1}^{K_w} w_k \rho^{-k}.$$
  If in addition $R_e >1$ and the covariance matrix of $I_{1:K_w}$ is
  strictly positive definite, then $\Cov(I_t, I_{t-j}) \rho^{j-2t}$
  converges to a positive constant
  for any fixed $j \geq 0$ and $t \rightarrow \infty$.
\end{lemma}

The proof is given in the Supplementary Material \ref{app-lemma1}. The second part of the
lemma implies that for $R_e > 1$
$E[I_t]^2/\Var(I_t)$ converges to a constant $>0$ as $t \rightarrow \infty$. In particular, the marginal distribution of $I_t$ differs strongly from a Poisson distribution. 

The infectivity profile $(w_k)$ can be approximated by the serial interval distribution \citep{cori2013new}. It is usually taken as fixed when estimating the reproduction numbers, or restricted to a
small number of proposed values. 

\begin{figure}[ht!]
	\begin{center}
		\includegraphics[width=0.5\linewidth]{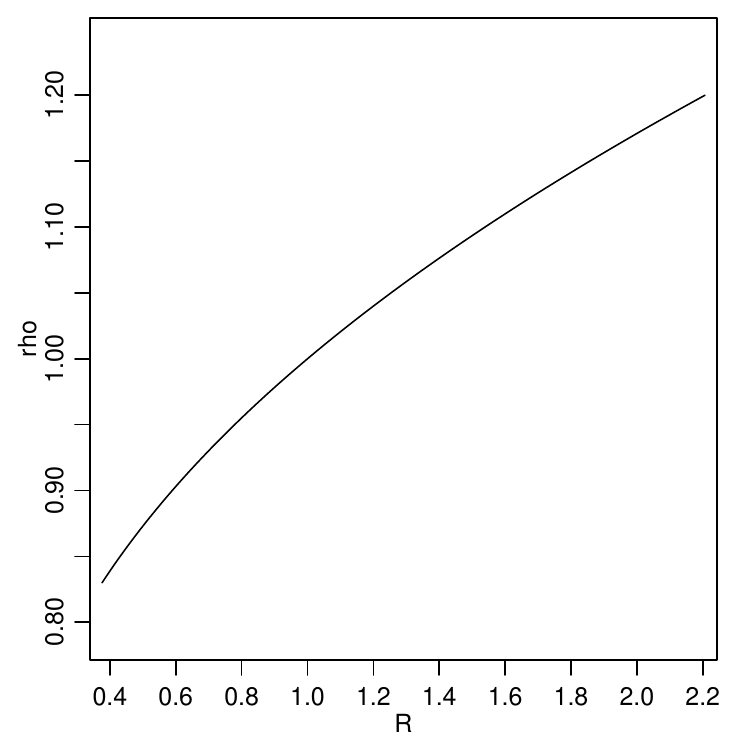}
		\caption{Asymptotic daily rate of change
                  $\rho =\lim_{t \rightarrow \infty}   E[I_{t+1})]/E[I_t]$ for a constant
                  reproduction number $R_e$ and the infectivity profile  of
                  \citet{huisman2022estimation}. Units on the $y$-axis are four times
                  the size of units on the $x$-axis.}
		\label{rateI}
	\end{center}
      \end{figure}

\subsection{Modeling disease detection}
The times between infection and detection are assumed to be independent
and identically distributed. The probability that an infection on day $s$ is
detected on day $t$ is denoted by $m_{s,t}$. In the simplest case, these
probabilities depend only on the difference $t-s$, i.e., $m_{s,t}=m_{t-s}$.
We assume that $m_{s,s}=0$ for all $s$ and that this
distribution has a time-independent upper limit $K_m$, i.e.  $m_{s,t}=0$ if
$t-s>K_m$. The probability $m_{s,+}$ that an infection on day $s$ is not
detected at all is
$m_{s,+}=1 - \sum_{k=1}^{K_m} m_{s,s+k}$. Undetected infections
are thus modeled with a defective delay distribution. For all $s$ and $t$ we have 
\begin{equation}
\label{eq:def-A}
I_s = \sum_{k=1}^{K_m} A_{s,s+k} + A_{s,+}, \quad D_t = \sum_{k=1}^{K_m}
A_{t-k,t}.
\end{equation}

By our assumptions, the variables
$(A_{s,s+1:K_m}, A_{s,+})$ and 
$(I_{s'}, A_{s',t}; s' \neq s)$ are conditionally independent given $I_s$ and 
\begin{equation}
  \label{eq:model-A}
  (A_{s,s+1:K_m}, A_{s,+}) \mid I_s \sim
  \mathcal{M}(I_s;m_{s,s+1}, \ldots, m_{s,s+K_m}, m_{s,+})
\end{equation}
where $\mathcal{M}$ denotes the multinomial distribution.

The delay is equal to the sum of the time from infection to onset of symptoms and the time from onset of symptoms to the time of case 
detection (confirmed test, hospital admission or death).
These two times are assumed to be independent. They are usually taken from the literature or estimated from a subset of data with known
onset of symptoms \citep{cori2013new} to obtain values $m_{s,t}$. Since case confirmation is usually affected by weekend and public holidays, these effects should either be removed in a preprocessing step or included in the delay distribution; see Subsection \ref{sub-weekday} and the Supplementary Material 
\ref{app-weekday}. The non-detection probabilities $m_{s,+}$ are
expected to depend on the number of tests during the days $s+1\!:s+\!K_m$, and one can try to estimate $m_{s,+}$ from this, see 
\cite{goldstein2024incorporating}. If the non-detection probability is constant in time, all values $I_s$ are underestimated by the same factor, and thus the estimated reproduction numbers are not affected.

\citet{abbott2020estimating} and \citet{flaxman2020estimating} assume that the detections $D_t$ are independent and have a negative binomial distribution with expectation $\sum _s m_{s,t} E[I_s]$ where the dynamics of $E[I_s]$ follows from \eqref{eq:model-I}.
In view of Lemma 1, this implies a much smaller stochastic variability compared to our model. Moreover, in our model the detections are conditionally dependent given the infection numbers.

\section {Methods}\label{methods}

We assume that the detected cases $D_t$ are available for days
$t \in 1\!:\!T$ and 
denote the data vector $D_{1:T}$ by $\mb D$. We first consider $T$ to be
fixed and discuss the case when observations become available sequentially in Subsection \ref{sub-seq} below.
The variables $I_s$, $A_{s,t}$, and $A_{s,+}$ are all unobserved,
and the likelihood of the reproduction numbers $R_{e,1:(T-1)}$ 
based only on the detections $D_{1:T}$ is intractable. It would be known explicitly if the infections $I_{(1-K_w):(T-1)}$ were available. In situations with latent variables, the EM algorithm is often used. In order to apply it directly to estimate 
$R_{e,1:(T-1)}$ from $D_{1:T}$, the conditional expectation of $I_s$ given $D_{1:T}$ should be tractable. But in our model this is not the case. 

\citet{goldstein2009reconstructing} use the EM algorithm to estimate the incidences $I_s$, not the reproduction numbers. The reproduction numbers $R_{e,s}$ are then estimated in a second step, substituting the estimates $\widehat{I}_s$ for $I_s$. Their EM algorithm is based on the assumption that given $(I_s)$, the latent variables $A_{s,t}$ are independent Poisson variables with parameters $I_s m_{s,t}$ which contradicts \eqref{eq:model-A}. Moreover, the dynamics of infections can be taken into account only through the starting value, and the choice of the starting value substantially influences the estimated incidences and reproduction numbers near the end of the observation period. See the Supplementary Material \ref{app-em} for more details on the EM algorithm to estimate $I_s$ and for an example of the sensitivity to the starting value.

We take a Bayesian approach instead, that is, we put a prior on the sequence of
reproduction numbers and approximate the joint posterior of the latent variables $I_s$ and $A_{s,t}$ and of the reproduction numbers $R_{e,t}$ given the observations by MCMC. This gives both 
point estimates and joint uncertainty quantification
of infections and reproduction numbers. Next, we describe this approach in more detail. 

 Since infections, which occurred during $K_m$ days preceding the data window, contribute to $\mb D$, we estimate $I_s$ also during these $K_m$
days and define $\mb I =I_{(1-K_m):(T-1)}$. We assume that the model
(\ref{eq:model-I})
for the dynamics of infections holds also during the additional $K_m$
days before the observations start. We then need the reproduction numbers $\mb R_e = R_{e, (1-K_m):(T-1)}$ and starting values
$\mb I_{init} = I_{(1-K_m - K_w):(-K_m)}$. We treat the latter also as unknown parameters
and put a prior on them. Finally, we set $\mb A = A_{(1-K_m):(T-1),1:T}$ where $A_{s,t} =0$ if $s \geq t$ or  $s <t-K_m$. The same convention is also used for $m_{s,t}$.

Conditionally on the unknown reproduction numbers $\mb R_e$ and on the initial values $\mb I_{init}$, the density of
$(\mb I, \mb A, \mb D)$ is given by
\begin{eqnarray}
p(\mb I,\mb A,\mb D \mid \mb R_e, \mb I_{init}) &=& p(\mb I \mid \mb R_e, \mb I_{init})p(\mb A \mid \mb I) p(\mb D \mid \mb A),
       \label{eq:fact-p}\\
  p(\mb I\mid \mb R_e, \mb I_{init})
  &=& \prod_{s=1-K_m}^{T-1} \frac{\exp(-\lambda_s +I_s\log \lambda_s)}{I_s!}, \label{p-I_R}\\
  p(\mb A \mid \mb I)
  &=& \prod_{s=1-K_m}^{T-1} I_s! \frac{(1 - b_s)^{I_s - B_s}}{(I_s - B_s)!}
      \prod_{t=\max(1,s+1)}^{\min(T,s+K_m)} \frac{m_{s,t}^{A_{s,t}}}{A_{s,t}!},  \label{p-A_I}\\
  p(\mb D \mid \mb A)
  &=& \prod_{t=1}^T \delta\left(D_t,\sum_{k=1}^{K_m} A_{t-k,t}\right).
  \label{p-D-A}
\end{eqnarray}
Here, $\lambda_s$ is defined in \eqref{eq:model-I},  $\delta$ is the Kronecker delta: $\delta(x,x)=1$
and $\delta(x,y)=0$ if $y \neq x$, and  
$$ B_s = \sum_{t=1}^T A_{s,t}, \quad b_s = \sum_{t=1}^T m_{s,t}.$$
Since  $B_s$ is the number of infections on day $s$ detected in the
window $1\!:\!T$,
$$I_s = B_s + \sum_{t=s+1}^0 A_{s,t} + 
    \sum_{t=T+1}^{s+K_m} A_{s,t} + A_{s,+}.$$
Thus \eqref{p-A_I} follows because by \eqref{eq:model-A}
$$
  (A_{s,1}, \ldots, A_{s,T}, I_s - B_s) \mid I_s \sim
  \mathcal{M}(I_s; m_{s,1}, \ldots, m_{s,T}, 1-b_s).
$$

The initial values $\mb I_{init}$ are assumed to be a priori 
independent and Poisson($\lambda^0_s)$-distributed:
\begin{equation}
    \label{eq:prior-I}
    p(\mb I_{init}) = \prod_{s=1-K_m-K_w}^{-K_m}
    \exp( -\lambda^0_s) (\lambda^0_s)^{I_s} /I_s!
\end{equation}
The prior for the effective reproduction numbers is chosen to be a Gaussian random walk 
on the log scale. With $L_t=\log R_{e,t}$ this means
$L_{1-K_m} \sim \N(0,\sigma^2)$ and the increments $L_{s+1}-L_{s}$
are independent and normal with mean 0 and variance $\tau^2$. 
Denoting the vector $L_{(1-K_m):(T-1)}$ by $\mb L$, we have 
\begin{equation}
    \label{eq:prior-L}
    p(\mb L) \propto \exp(\left( - \frac{1}{2 \sigma^2} L_{1-K_m}^2 -  \frac{1}{2\tau^2} \sum_{s=1-K_m}^{T-2} (L_{s+1}-L_s)^2\right).
\end{equation}
We can easily adapt our method to other Gaussian priors for $\mb L$.

The hyperparameters in our approach are 
therefore $\lambda^0_{(1-K_w-K_m ):(-K_m)}$ and
$(\sigma^2$, $\tau^2)$. As the values $\lambda^0_s$ only provide initial
conditions for  $I_s$ on days $s$ that are well separated from
the observation window, we expect  that the prior for
$\lambda^0_s$ does not have a strong influence on the posterior as long as
there is no clear conflict with the data.
We discuss our choice of this prior
in the Supplementary Material \ref{app-prior-mean-I}. The prior standard deviation
$\sigma$ of $L_{1-K_m}$ will be chosen fairly large, e.g., between
1.5 and 2, in order to be only weakly informative. In contrast, the
parameter $\tau$ must reflect our belief how quickly the reproduction
number can change from one day to the next. Typically, it will be small, e.g., 0.025. Roughly speaking, this means that, a priori, we
expect the daily changes of $R_{e,t}$ to be around $2.5\%$ on average
and rarely larger than $5\%$. The influence of $\tau$ on the results in our data example is assessed in Subsection \ref{sub-sub-prior-L} by means of a sensitivity study. One could
also let $\tau$ depend on time to allow larger changes when the measures taken by
the government against the spread of the epidemic change, or one could
let $\tau$ depend on the reproduction number of the previous day if a
constant relative change seems not appropriate.

The posterior $p(\mb I,\mb A, \mb L, \mb I_{init} \mid \mb D)$ is proportional to the product of the 
priors given in \eqref{eq:prior-I} and \eqref{eq:prior-L} times the likelihood given in \eqref{eq:fact-p} - \eqref{p-D-A}. In order to analyze this posterior, we draw samples from it using 
a component-wise MCMC algorithm that alternates between sampling from
$p(\mb I,\mb A, \mb I_{init} \mid \mb L,\mb D)$  and 
$p(\mb L \mid \mb I, \mb I_{init},\mb A, \mb D)=p(\mb L \mid \mb I, \mb I_{init})$.
These two sampling algorithms are introduced in the next two subsections.
The choice of the starting value for the MCMC algorithm is described in the Supplementary Material \ref{app-init-val}.

If $T \geq \max(K_m, K_w)$, we can easily obtain also samples of posterior
predictive distributions
from the posterior samples. This is due to the fact that the
process $(\mb x_t)$ is a Markov chain if we define
\begin{equation}
  \label{eq:state}
  \mb x_t = (L_{t-1}, I_{(t-K_w):(t-1)}, A_{(t-K_m):(t-1),t},
U_{(t+1-K_m):(t-1),t})
\end{equation}
where
$$U_{s,t} = \sum_{s'=t+1}^{s+K_m} A_{s,s'} + A_{s,+} $$
is the number of infections on day $s$ not detected until day $t$.
The transition density of this chain is
\begin{eqnarray}
p(\mb x_{t+1} \mid \mb x_t) &=& p_{norm}(L_t; L_{t-1}, \tau^2) \cdot
p_{Poisson}(I_t; \lambda_t) \cdot p_{binom}(A_{t,t+1}; I_t, m_{t,t+1})
                                \nonumber\\
&\cdot& \prod_{s=t+1-K_m}^{t-1} p_{binom}\left(A_{s,t+1}; U_{s,t},
        \frac{m_{s,t+1}}{m_{s,t+1} + ... + m_{s,s+K_m} + m_{s,+}}\right)
        \nonumber \\
&\cdot& \delta(U_{t,t+1},I_t- A_{t,t+1})\prod_{s=t+2-K_m}^{t-1}
        \delta(U_{s,t+1},U_{s,t}-A_{s,t+1}).   \label{eq:state-trans}
\end{eqnarray}
Sampling from the transition density for given $\mb x_t$ is easy, and a
sample from the posterior contains a sample from $p(\mb x_T \mid \mb D)$ if
$T \geq K_m$ because $U_{s,T} = I_s - B_s$ if $s \geq 0$.

\subsection[Sampling]{Sampling from $p(\mb I,\mb A,  \mb I_{init} \mid \mb L,\mb D)$}
\label{sub-sample-I}

The conditional distribution $p(\mb I, \mb A, \mb I_{init} \mid \mb L, \mb D)$ is the most difficult to sample from. Given the current values  $(\mb I, \mb A, \mb L)$, we use an independent Metropolis-Hastings
step that proposes new values $(\mb I^*, \mb A^*, \mb I^*_{init})$ as follows. The variables $\mb A^*$ are generated by drawing the columns $A^*_{(t-K_m):(t-1),t}$ for $t \in 1\!:\!T$ from independent multinomial distributions with size $D_t$ and probabilities
  $$\nu^*_{s,t}= \frac{\psi^*_s m_{s,t}}{\pi^*_t}, \quad
  \pi^*_t = \sum_{s=t-K.m}^{t-1}\psi^*_{s} m_{s,t}.$$
  The $\psi^*_s$ are arbitrary
  positive values which can depend on $\mb L$ and 
  $\mb D$. The goal is to choose them
  such that the acceptance probability is high. We first initialize $\psi^*_s = \lambda^0_s$ for 
  $1-K_m-K_w \leq s \leq -K_m$ and iterate for $s \in (1-K_m)\!:\!(T-1)$
  $$
  \psi^*_s = \exp(L_s) \sum_{k=1}^{K_w} w_k \psi^*_{s-k}. 
  $$
  Finally, we apply one EM iteration as described in the Supplementary Material \ref{app-em} to $\psi^*_{(1-K_m):(T-1)}$, using the observations $\mb D$.
  The intuition for this choice is discussed in Subsection \ref{sub-sub-psi-star} below.

  The variables $B^*_s$ for $s \in (1-K_m)\!:\!(T-1)$ are determined by $\mb A^*$. It remains to propose 
  $\mb I^*_{init}$ and the variables
  $I_s^* - B_s^*$ for $s \in (1-K_m)\!:\!(T-1)$. The initial values are generated according to our prior \eqref{eq:prior-I}.
  For $s \in (1-K_m)\!:\!(T-1)$, we then compute recursively
  $$\lambda_s^* = \exp(L_s) \sum_{k=1}^{K_w} w_k I_{s-k}^*$$
  and generate
  $$ I_s^* - B^*_s \sim \textrm{Poisson}((1-b_s)\lambda_s^*). $$
  
The proposal is concentrated on the set
$\{\sum_{k=1}^{K.m} A^*_{t-k,t} =D_t; t \in 1\!:\!T\}$ and its density has
the closed form expression 
\begin{eqnarray}
  \label{eq:fact-q1}
    q(\mb I^*,\mb A^*, \mb I^*_{init} \mid \mb L, \mb D)
  &=& \prod_{t=1}^T D_t! \prod_{s=t-K_m}^{t-1}
  \frac{(\nu^*_{s,t})^{A^*_{s,t}}} {A^*_{s,t}!} \cdot
      \prod_{s=1-K_m-K_w}^{-K_m} \exp(- \lambda^0_s)
      \frac{(\lambda^0_s)^{I^*_s}}{I^*_s!} \nonumber \\
      &\cdot& \prod_{s=1-K_m}^{T-1} \exp(- (1-b_s)\lambda^*_s)
  \frac{((1-b_s)\lambda^*_s)^{I^*_s-B^*_s}}{(I^*_s-B^*_s)!}.
  \end{eqnarray}
  

Since $p(\mb D \mid \mb A) = p(\mb D \mid \mb A^*) = 1 $, the acceptance
probability is given by 
 $\min(1, r)$ where
$$r = r(\mb I^*, \mb A^*, \mb I^*_{init}, \mb I, \mb A, \mb I_{init}, \mb L, \mb D) = 
\frac{p(\mb I^*_{init})p(\mb I^* \mid \mb L, \mb I^*_{init}) p(\mb A^* \mid \mb I^*)
  q(\mb I,\mb A, \mb I_{init} \mid \mb L, \mb D)}
{p(\mb I_{init})p(\mb I \mid \mb L, \mb I_{init}) p(\mb A \mid \mb I)
  q(\mb I^*,\mb A^*, \mb I^*_{init} \mid \mb L, \mb D)}.$$

The terms on the right-hand side are given in equations
\eqref{p-I_R}, \eqref{p-A_I}, \eqref{eq:prior-I} and 
\eqref{eq:fact-q1}. With some algebraic manipulations detailed in the Supplementary Material \ref{app-acc-ratio},
we obtain the following explicit form
\begin{eqnarray}
  \label{eq:prop1-r}
  \log r(\mb I^*, \mb A^*, \mb I^*_{init}, \mb I, \mb A, \mb I_{init}, \mb L, \mb D)
  &=& \sum_{s=1-K_m}^{T-1}\left(B^*_s \log\frac{\lambda_s^*}{\psi^*_s} -
      B_s \log \frac{\lambda_s}{\psi_s} -
      b_s(\lambda^*_s - \lambda_s)\right)  \nonumber \\
  & & + \sum_{t=1}^T D_t \log \frac{\pi^*_t}{\pi_t}.
\end{eqnarray}
In particular, $r$ depends on $\mb A$ and $\mb A^*$ only through $\mb B$ and
$\mb B^*$.

\subsubsection{Choice of the parameters $\psi^*_s$ in the proposal
  of $\mb I^*$}
\label{sub-sub-psi-star}

Ideally, we would like to choose the values $\psi^*_s$ such that they
satisfy at the same time
$$
  \pi^*_t = \sum_s \psi^*_s m_{s,t} = D_t \, \, (1 \leq t < T) 
  \textrm{ and }
  \psi^*_s = \exp(L_s) \sum_k w_k \psi^*_{s-k} \, \, (s \geq 1-K_m).
$$
The first condition implies that $E_q[A^*_{s,t}]= \psi^*_s m_{s,t}$ and $E_q(B^*_s) = \psi^*_s b_s$. Together with the second condition, this implies that for $s> -K_m$
$$E_q[I^*_s] = \psi^*_s + (1- b_s)(E_q[\lambda^*_s] - \psi^*_s) = \psi^*_s + 
(1-b_s)\exp(L_s) \sum_k w_k (E_q[I^*_{s-k}] - \psi^*_{s-k}).$$
For $s \leq -K_m$ it holds by construction that 
$E_q[I^*_s] = \psi^*_s$ and therefore they are equal for  all $s$. This
means that under the proposal $I^*_s$ is compatible with the observations and satisfies the dynamics of the model on average.

%

\subsection{Sampling from $p(\mb L \mid \mb I, \mb I_{init})$}
\label{sub-sample-L}

The target density $p(\mb L \mid \mb I, \mb I_{init})$ is proportional to
$$ p(\mb L) p(\mb I \mid \mb L, \mb I_{init}) \propto
\exp\left(- \frac{L_{1-K_m}^2}{2 \sigma^2} - \sum_{t=1-K_m}^{T-2} \frac{(L_{t+1}-L_t)^2}
{2 \tau^2} + \sum_{t=1-K_m}^{T-1} (I_t(L_t+\log \kappa_t) - e^{L_t} \kappa_t)\right).$$
where $\kappa_t=\sum_{k=1}^{K_w} w_k I_{t-k}$. Therefore, up to additive terms
that do not contain $\mb L$,  
$$\log p(\mb L \mid \mb I, \mb I_{init})=- \frac{L_{1-K_m}^2}{2 \sigma^2} -
\sum_{t=1-K_m}^{T-2} \frac{(L_{t+1}-L_t)^2}
{2 \tau^2} + \sum_{t=1-K_m}^{T-1} (I_tL_t - e^{L_t} \kappa_t).$$
Except for the terms $e^{L_t}$ this is a quadratic form in $\mb L$.
We thus use a Metropolis-Hastings update with a Gaussian proposal
for $\mb L^*$. The mean and covariance of the proposal and the computation of the acceptance ratio are described in the Supplementary Material \ref{app-prop-Lstar}.

\subsection{Sequential estimation}\label{sub-seq}

If each day a new observation becomes available, the observation window
and therefore also the dimension of $(\mb L, \mb I, \mb A)$ increases.
This means that sampling from the posterior becomes more difficult as more
observations become available. To reduce the computational burden, one would
like to take advantage of the already available samples since typically
the posterior of variables in the distant past changes little when we condition also on this new observation.

For such situations sequential Monte Carlo (SMC) methods have been
developed, see e.g. \citet{chopin2020introduction}. SMC methods are
especially suitable for models that can be put in the state space
framework, that is if the observation $D_t$ is a function of the state at
time $t$ of a latent Markov process $(\mb x_t)$. This is satisfied if the
state process is defined as in \eqref{eq:state} and \eqref{eq:state-trans}.

Unfortunately, our attempts to use SMC methods in our case failed. The main
reason is that a new observation affects variables up to two weeks in time
backward. Thus both the bootstrap and the auxiliary particle filter
collapse easily since they modify none of the already available state variables at all or only the most recent one.
\citet{gilks2001following} have proposed rejuvenation steps
to modify all or many backward variables. In our case, the proposals for
rejuvenation of infections have a low acceptance rate so that the gains in
computational effort are minimal. Similarly, in the block sampling method
of \citet{doucet2006efficient}, we were not able to devise
an approximation of the optimal proposal distribution in their equation
(3.9) that avoids sample collapse. 

We therefore applied a simple sequential version which computes posteriors
given moving windows $D_{(t-\ell+1):t}$ of fixed length $\ell$ for
$t=\ell, \ell+1, \ldots$ by the MCMC algorithm described in Subsections
\ref{sub-sample-I} and \ref{sub-sample-L}.
When a new observation $D_t$ becomes available, we
use the samples from the posterior given
$D_{(t-\ell):(t-1)}$ only to choose the prior
of the initial values $I_{(t-\ell+1 - K_m -K_w):(t-\ell - K_m)}$. Apart from this,
we start from scratch and ignore all previous results. We have tried
to use the sample given $D_{(t-\ell):(t-1)}$ also for the starting
value of the MCMC algorithm, but found that in some cases, this leads to
bad starting values, essentially for the same reason that caused sample
collapse of SMC methods. 

With this rolling window approach, the computational effort is independent of
$t$. The disadvantage is that for $s$ near the beginning $t-\ell +1$ of the
new data window, the estimated quantiles of $I_s$ and $R_{e,s} $ can differ
substantially from those already available from previous iterations. As the
latter use more observations close to $s$ and seem thus more reliable,
we recommend using the most recently estimated quantiles only for
$s \in (t-\ell/2)\!:\!(t-1)$ and otherwise keep those from the previous
iteration. If smooth quantiles are desired, one can use interpolation of estimated
quantiles in a small transition  period. In our examples, we  will choose the window length equal to $42$, i.e., six weeks.
In the simulation example of Subsection \ref{simulation}, the wall-clock time for analyzing one such window was 295 seconds with our MCMC algorithm and 59 seconds with the method of \citet{huisman2022estimation} on a laptop with an
Intel i7-12800H processor and 32 GB of random-access memory using a single core.

\subsection{Weekday effects}
\label{sub-weekday}

\begin{figure}[ht!]
\begin{center}
 \includegraphics[width=\textwidth]{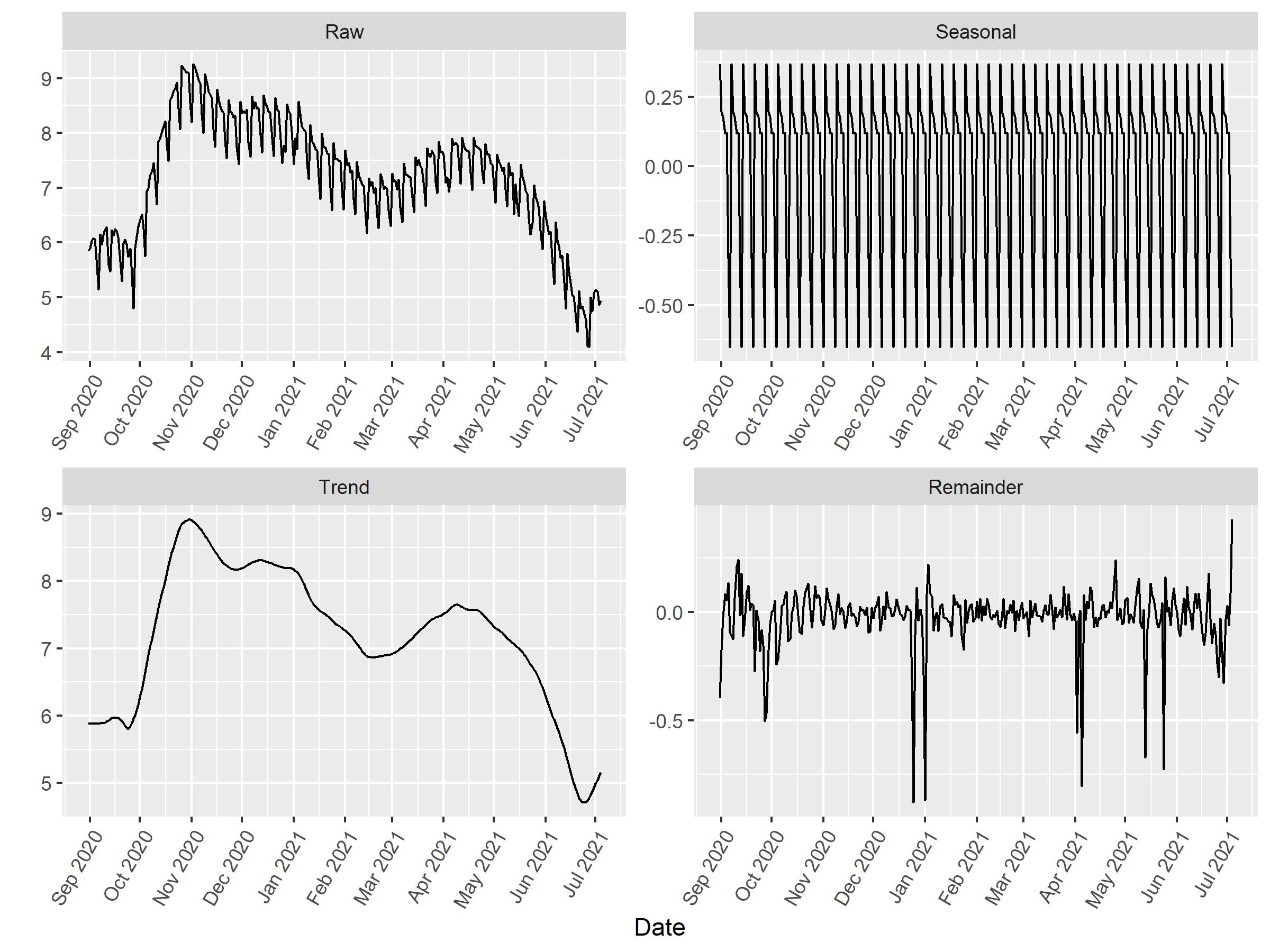}
 \caption{Logarithm of daily positive test results in Switzerland from
  August 31, 2020, until July 4, 2021 (top left), together with a
  decomposition into weekday effect (top right), trend (bottom left)
  and remainder (bottom right). The data were downloaded on July 14, 2021,
  and the decomposition was computed with the function \texttt{stl} in $R$,
  using the parameters t.window=15, s.window=7, robust=TRUE.}
\label{stl-1}
\end{center}
\end{figure}

An additional difficulty arises from the fact that in practice the
observations $D_t$ contain strong weekday and holiday effects. The top-left plot
of Fig. \ref{stl-1}
shows the time series of log detections in Switzerland during a ten  month
period.  A weekday effect is
clearly visible. As it is approximately constant on the logarithmic scale,
the weekday effect is multiplicative.  The figure also shows the smooth
trend $\bar{D}_t$, weekday effects and remainder obtained from the additive
decomposition of $\log D_t$ by the method of \citet{rb1990stl}, using the
function \verb+stl+ in $R$.
Holiday effects appear as outliers in the remainder since we use the robust
version of the method.  The effect of holidays which did not fall on a
weekend (Christmas Eve, Christmas, New Year's Eve, New Year's Day, Good Friday,
Easter Monday, Ascension and Whitsun Monday) is clearly visible.
The cause of somewhat smaller outliers at the end  of September 2020 and at the
end of June 2021 is not clear.

In order to deal with weekday effects, one should choose a delay
distribution that depends on the day of the week. In the Supplementary Material \ref{app-weekday}, we describe two ways for introducing a weekday effect
into a time invariant delay distribution. If we assume these weekday
effects to be time-invariant, we could choose a prior for them and estimate
them at the same time as infections and reproduction numbers.  However, 
Figure \ref{weekdays} in the Supplementary Material indicates that the weekday effects also change slowly
over time. Moreover, we would then also have to estimate a delay
distribution which can handle holiday effects  and 
other artifacts visible in the remainder term in Fig.
\ref{stl-1}. Finally, since the reproduction numbers depend on contact
rates, weekday and holiday effects could also
be present in the reproduction numbers.

We leave this problem for future research and use, following \citet{huisman2022estimation}, a simpler procedure: We first remove weekday effects and
estimate smooth trend data $\bar{D}_{1:T}$ from raw data $D_{1:T}$ and then draw samples from the posterior given $\bar{D}_{1:T}$ instead of
$D_{1:T}$, assuming a time-invariant delay distribution. 

In the sequential version, we use this procedure for each data window
$D_{(t-\ell+1):t}$. A smooth trend $\bar{D}_{(t-\ell+1):t}$ is estimated
from $D_{(t-\ell+1):t}$ and 
then we sample from the posterior of infections and reproduction
numbers given $\bar{D}_{(t-\ell+1):t}$. This means
that the data $\bar{D}_{(t-\ell+1):t}$ on which we condition differ
somewhat in overlapping parts of two windows.
We will investigate how big this effect is in the examples in
Subsection \ref{sub-short}.

For estimating the smooth trend and for removing the weekday effects, we use
the method of \citet{rb1990stl} which is implemented in $R$ by the
function \verb+stl+. Since the weekday effects are multiplicative, we use
it on the log scale. It has two tuning parameters
t.window and s.window which determine the amount of smoothing for the trend and the seasonal (in our case weekday)
component. The larger these parameters, the stronger the smoothing.
We chose the values t.window=15 and s.window=7 for large observation
windows and t.window=15 and s.window=''periodic'' (i.e. maximal smoothing)
for the windows of length 42 in the sequential implementation. We also
make a multiplicative adjustment such that the sums of $D_t$ and
$\bar{D}_t$ agree. \citet{huisman2022estimation} use
a different smoothing method, but otherwise they deal with weekday effects in the same way. In our experiments in Subsection \ref{sub-long}, we will also compare the two smoothing approaches. 

\section{Extensions of model and methods}
\label{extension}
The model described in Section \ref{sec_notation} is relatively simple. We present here two generalizations that increase the flexibility without adding too much complexity.

The first one replaces the conditional mean
$\lambda_t$ in \eqref{eq:model-I} by a random mean
which is Gamma-distributed with shape $\gamma$ and 
expectation $\lambda_t$.
Then the rate parameter is $\eta_t = \frac{\gamma}{\lambda_t}$.
It is well-known that under this assumption
\eqref{eq:model-I} is replaced by 
\begin{equation}
  \label{eq:model-I-ext}
  I_t \mid (I_s; s<t) \sim
  \textrm{Neg-Binom}\left(\gamma, \frac{\gamma}{\gamma + \lambda_t}\right).
\end{equation}
where Neg-Binom denotes the negative binomial distribution.
In particular, the conditional variance of $I_t$ is then
$\lambda_t + \lambda_t^2/\gamma$ instead of $\lambda_t$. 
Introducing such overdispersion has been suggested repeatedly in the literature \citep[e.g.,][]{lloyd2005superspreading, mishra2022covid, bhatt2023semi} since it can take emigration and immigration of infectious individuals and superspreaders into account.

Our MCMC algorithm can be adapted to the modified target distribution by changes at two places. For updating $\mb A$ and $\mb I$ given the reproduction numbers, one should generate the variables
$U^*_s = I_s^* - B_s^*$ for $s \in (1-K_m)\!:\!(T-1)$ not from Poisson
distributions with  parameter $(1-b_s)\lambda_s^*$, but from the conditional distributions given $B^*_s$ and $\lambda^*_s$ under \eqref{eq:model-I-ext}:
$$ U^*_s \sim \textrm{Neg-Binom}\left(B^*_s + \gamma, 
   \frac{b_s\lambda^*_s + \gamma}{\lambda^*_s + \gamma}\right).$$
The derivation of this together with the modified formula for
the acceptance probability is given in the Supplementary Material \ref{app-ext}. 

For updating $\mb R$ given $\mb I$, we have to compute the second order Taylor approximation of $\log p(I_t \mid \lambda^*_t)$ at the current value $R_t$ for the negative binomial instead of the Poisson density. The formulae for the mean and the precision matrix of the Gaussian proposal distribution are given in the Supplementary Material \ref{app-ext}.

The second possible extension of the model uses a random delay distribution with a Dirichlet prior centered at 
$(m_{s,s+1:K_m}, m_{s,+})$. Given $I_s$, the variables $A_{s,s+1:K_m}, A_{s,+}$ have then a Dirichlet-multinomial distribution. Again, this leads to overdispersion which can be justified by immigration or emigration of infected, but not yet detected cases. Adaptation of our MCMC algorithm to this modified distribution seems more difficult and is left open for further work.

\section{Applications to observed and simulated data}\label{examples}

We first apply our method to the data of daily positive test results in Switzerland in the period from August 31, 2020 until July 4, 2021, and compare our results with those obtained by the method of \citet{huisman2022estimation}. The data can be downloaded from the website \url{https://www.covid19.admin.ch/de/epidemiologic/case}. August 31, 2020, which was a Monday, is our day one, and Sunday July 4, 2021, is our day number 308. The code, which implements our methods and for doing the following experiments, can be found at \if1\blind{\url{https://github.com/fabsig/BayesianReproductionNumber}}\else{\#anonymized link\# }\fi.

In Section \ref{simulation}, we also run a simulation experiment where we take the reproduction numbers from August 3 to October 31, 2020, estimated from the actual data as the truth and then generate samples according to our model. Furthermore, we look in more detail at the sequential estimates with data from August 31 to November 8, 2020, in Section \ref{sub-short}.

For comparison with the method of \citet{huisman2022estimation} we adopt all their choices, except that we smooth the data with the approach of \citet{rb1990stl} described in Subsection \ref{sub-weekday}. Specifically, the values $w_k$ for the infectivity
profile are obtained by discretizing and truncating the Gamma-distribution with mean 4.8 and standard deviation 2.3 to integers $1, \ldots, K_w=12$. Numerical values are given in the Supplementary Material \ref{app-delay-comp}. Fig. \ref{rateI} shows the growth rate  $\rho= \rho(R_e)$ of Lemma 1 for this choice. 
The delay distribution is assumed to be time-invariant 
with $m_+=0$ and $K_m=28$, obtained by
discretizing and truncating the convolution of two Gamma-distributions, corresponding to time from infection to onset of symptoms and to time from onset of symptoms to case confirmation. For the former, we choose mean 5.3 and standard deviation 3.2, for the latter mean 5.5 and standard deviation 3.8. 
See the Supplementary Material \ref{app-delay-comp} for the computation and the values of $m_k$.

\subsection{Results for the period August 31, 2020 until July 4, 2021}\label{sub-long}

\begin{figure}[ht!]
	\begin{center}
		\includegraphics[width=\textwidth]{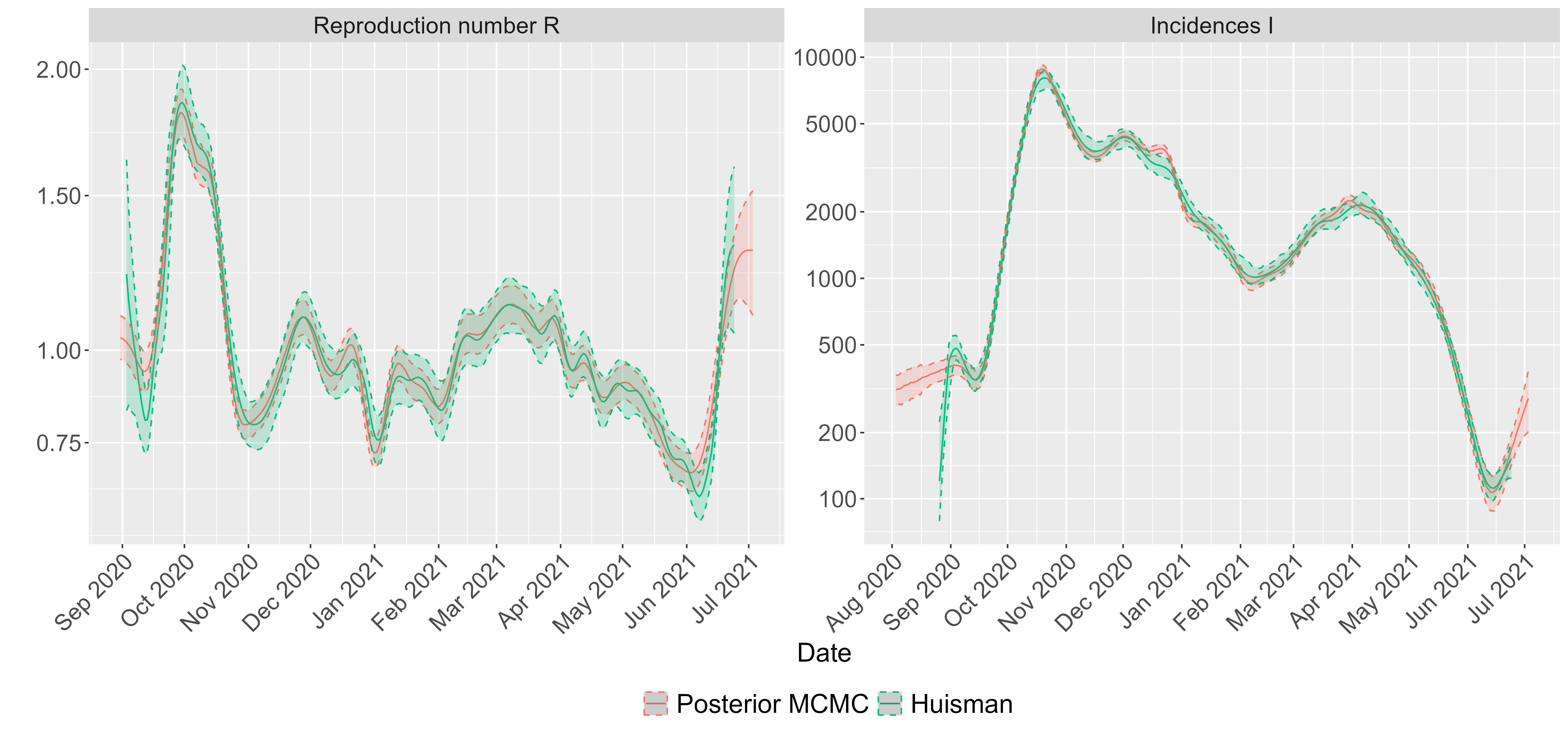}
		\caption{Posterior 2.5\%-, 50\%- and 97.5\%-quantiles of reproduction and infection numbers from August 3, 2020 until July 3, 2021. Estimates and confidence intervals according to the \citet{huisman2022estimation} method are also shown for comparison. }
		\label{post}
	\end{center}
\end{figure}

Fig. \ref{post} shows 2.5\%, 50\%, and 97.5\% posterior quantiles of $I_s$ and $R_{e,s}$ given the entire observation window of 308 days by our method and that of \citet{huisman2022estimation}.
For the parameters $\sigma$ and $\tau$ of the prior for $L$, we used the values $1.5$ and $0.025$, respectively, based on the discussion following the definition of the prior \eqref{eq:prior-L}. The prior means $\lambda^0_s$ in equation \eqref{eq:prior-I} were chosen as described in Section C.2 of the Supplementary Material. Convergence of our MCMC algorithm was monitored by inspecting trace plots.

Although the agreement in the point estimates of the two methods is good in general, there are several periods with discrepancies. The
differences among the two methods in both the location and width of
the confidence intervals are particularly large at the beginning and
end of the observation period. Note that the code of
\citet{huisman2022estimation}, see
\url{https://github.com/covid-19-Re/shiny-dailyRe/}
(as of June 19, 2025), produces estimates of $R$ only for days $4\leq t \leq T-10$ and of $I$ only for
days $-4\leq t \leq T-10$. Details are given in Supplementary Material
\ref{app_sub_implement}. In principle, 
their method could produce estimates for the same period as ours. We conjecture that the reason for this omission is that their
estimates are often very inaccurate at the end and start of the observation
period; see also the simulation study in Section \ref{simulation}. In practice, accurate inference for $R_e$ at the
end of an observation period is important for timely reporting. We thus provide estimates until the very end. 

 In addition to the most obvious differences at the start and end of the observation period, the results also differ at several places inside the observation period. For example,
our method estimates the start of a new increase in the reproduction
number at the end of October a few days earlier. Furthermore, there
are frequent differences in the width of the intervals, the confidence
intervals of \citet{huisman2022estimation} being often wider than ours. 

The incidences at the beginning of the observation window estimated by the method of \citet{huisman2022estimation} drop suddenly because their code applies zero padding to the data at the beginning. We believe that this is due to the fact that they intend to apply their method already in the initial phase of the epidemic.
As the dynamical model of the epidemic is 
based on a mean field approximation of contacts between infectious and susceptible individuals, it is likely that such a model does not fit well during the early phase of an epidemic. We thus chose an observation period which starts later.

\citet{huisman2022estimation} used in the preprocessing step local polynomial
regression (LOESS) whereas we used the method of
\citet{rb1990stl} described in Section 
\ref{sub-weekday}. LOESS does not estimate weekday effects, but removes them by smoothing. 
In order to see whether this difference matters, we repeat the analysis
with the data preprocessed as in \citet{huisman2022estimation} for both methods. We use the same LOESS tuning parameters as \citet{huisman2022estimation}: the degree of the polynomials is one and the smoothing span is $21/T$. The results are shown in Fig. \ref{post_loess} in Supplementary Material \ref{app_sub-long}. We again find that there are differences between results from our MCMC sampler and the method of \citet{huisman2022estimation}. Apart from the strong differences at the end of the observation period, the bootstrap confidence intervals of \citet{huisman2022estimation} show relatively large oscillations.

\subsubsection{Influence of the prior for $\mb L$}\label{sub-sub-prior-L}
\begin{figure}[ht!]
  \begin{center}
    \includegraphics[width=\textwidth]{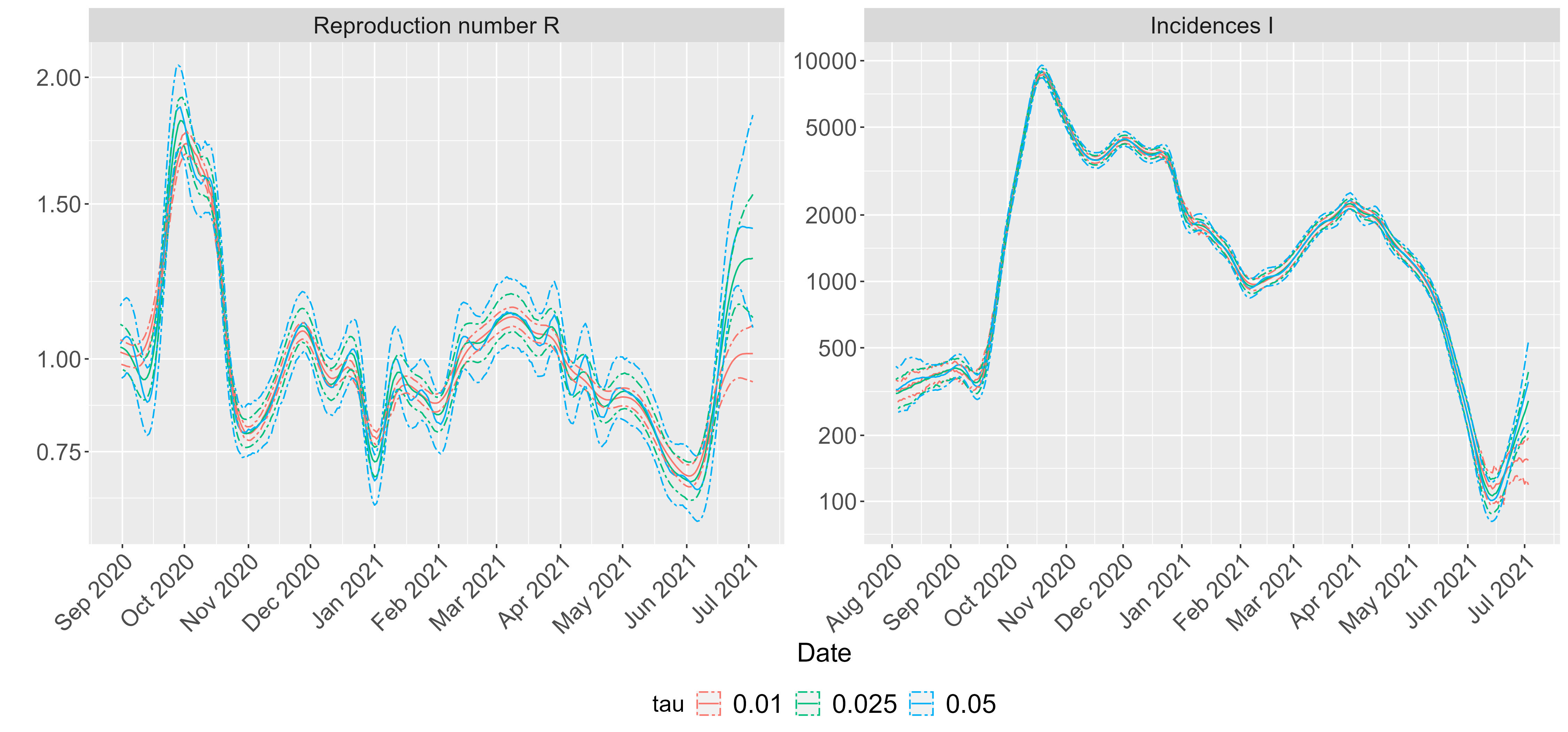}
    \caption{Posterior 2.5\%-, 50\%- and 97.5\%-quantiles 
      for different choices of $\tau$. }
    \label{sensitivity_tau}
  \end{center}
\end{figure}

  In order to assess the effect of varying the value of $\tau$ we additionally inspect the posterior distributions for $\tau=0.01$ and $\tau=0.05$ instead of $\tau=0.025$, with the other specifications unchanged. Fig. \ref{sensitivity_tau} shows that the posterior distributions exhibit quite some sensitivity with respect to the choice of $\tau$. With increasing $\tau$ the posterior becomes wider
and there is more oscillation. Near the beginning and end of the period
also the location of the posterior depends on $\tau$. This is to be
expected as there is little information in the data about how much the
reproduction number can change in a day. In order to take the uncertainty
about $\tau$ into account, one could use a hierarchical Bayesian approach and specify a hyperprior for distribution for $\tau$ instead of a fixed value. We leave the investigation of these ideas for further research. 

\subsection{Simulation experiment}\label{simulation}

\begin{figure}[ht!]
  \begin{center}
    \includegraphics[width=0.49\textwidth]{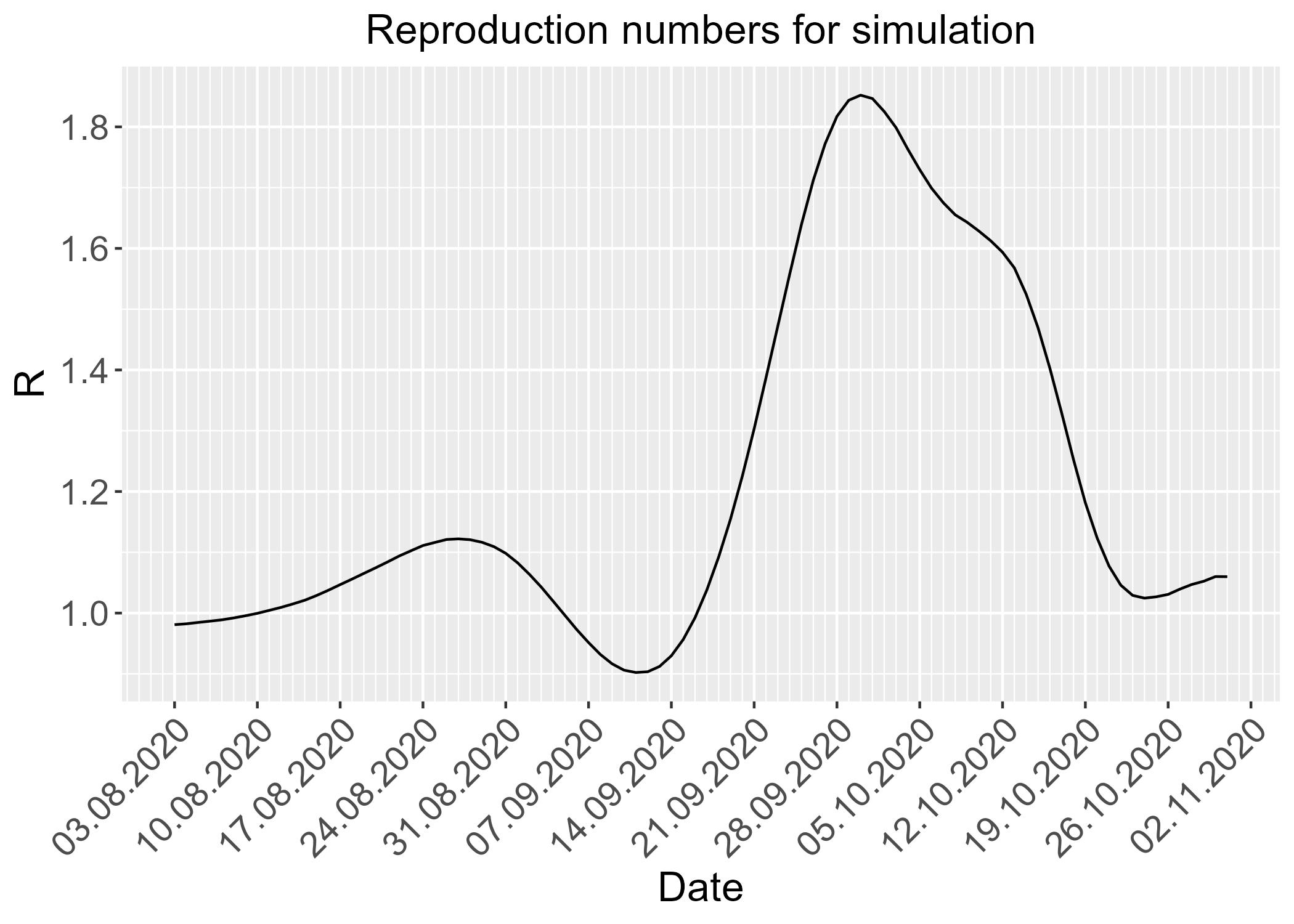}
    \includegraphics[width=0.49\textwidth]{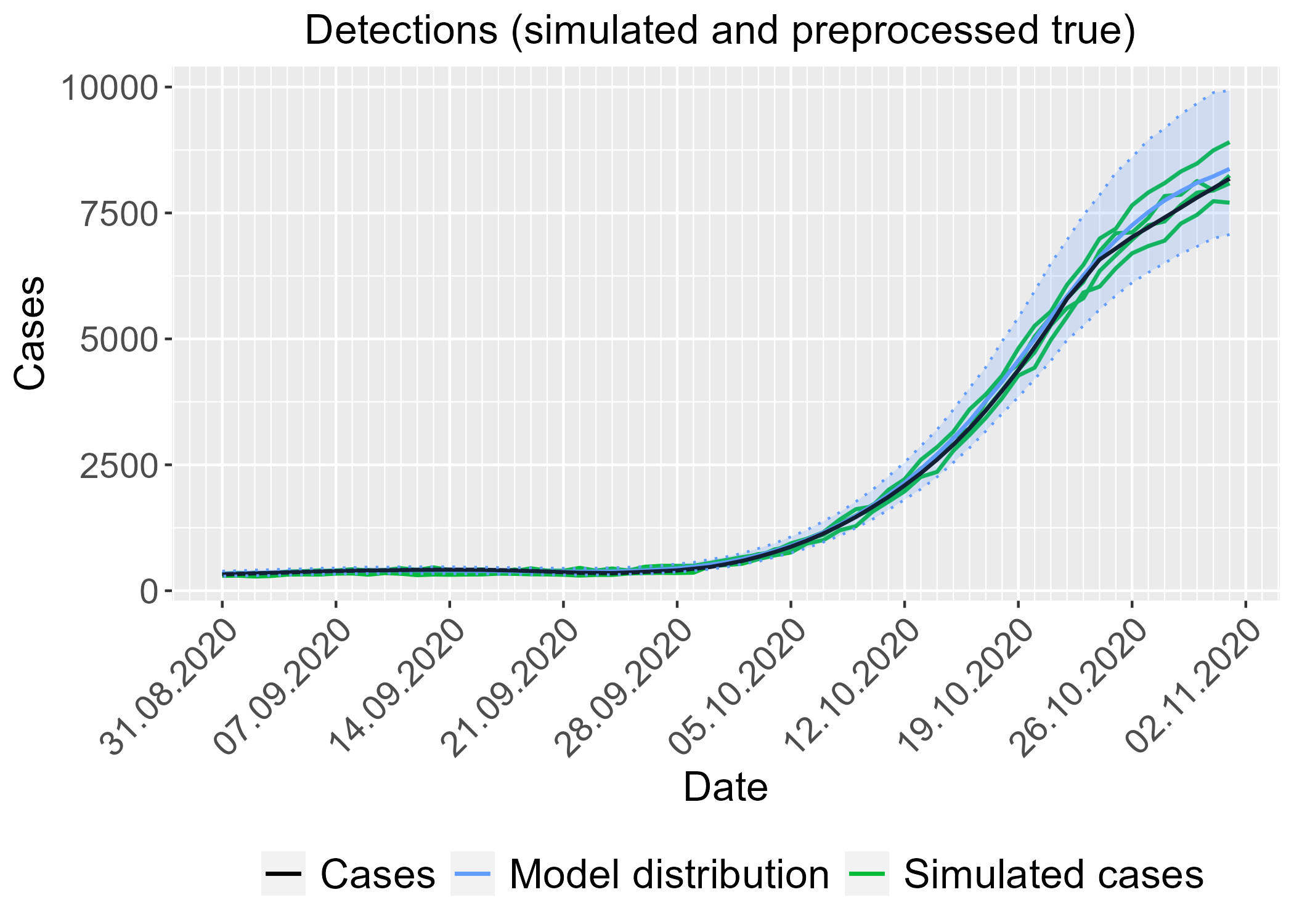}
    \caption{Left: Posterior median of reproduction numbers from August 3 to October 31 given detections from August 31 to November 1, used as ground truth in the simulation study.
    Right: Samples and 2.5\%, 50\% and 97.5\% quantiles of simulated detections (color) and observed detections from August 31 to November 1 after preprocessing (black).}
    \label{R_simulation_study}
  \end{center}
\end{figure}

In the following, we conduct a simulation experiment to assess the
accuracy of our method and compare it with the approach of \citet{huisman2022estimation}. In order to run multiple simulation iterations in a reasonable amount of time, we use a shorter observation period 
and simulate detection data $D$ only for the first nine weeks
($63$ days) from August 31 to November 1, 2020. Simulation is done as follows. As ground truth for the reproduction numbers $R_{e,s}$, we choose the posterior median of $R_{e,s}$ for $s$ running from August 3 to October 31, estimated using data from August 31 to November 1, 2020, with the same specifications as in Subsection \ref{sub-long}. This ground truth is shown in the left plot of Fig. \ref{R_simulation_study}. Infection and detection numbers are then simulated $100$ times using \eqref{eq:prior-I} for $\mb I_{init}$ 
and \eqref{p-I_R} -- \eqref{p-D-A} for $\mb I$, $\mb A$ and $\mb D$. The values $\lambda^0_s$ in \eqref{eq:prior-I} were the same as those used for the results in Subsection \ref{sub-long}. As there are no weekday effects, we do not use any preprocessing for our method. For the method of \citet{huisman2022estimation}, we had to use preprocessing because otherwise the results are much worse due to zero padding and their block bootstrap. Specifically, we use the \verb+stl+ preprocessing function \citep{rb1990stl}. The right plot of Fig. \ref{R_simulation_study} shows the data together with a few generated samples and the 2.5\%, 50\% and 97.5\% quantiles of the simulation model.

\begin{figure}[ht!]
  \begin{center}
    \includegraphics[width=\textwidth]{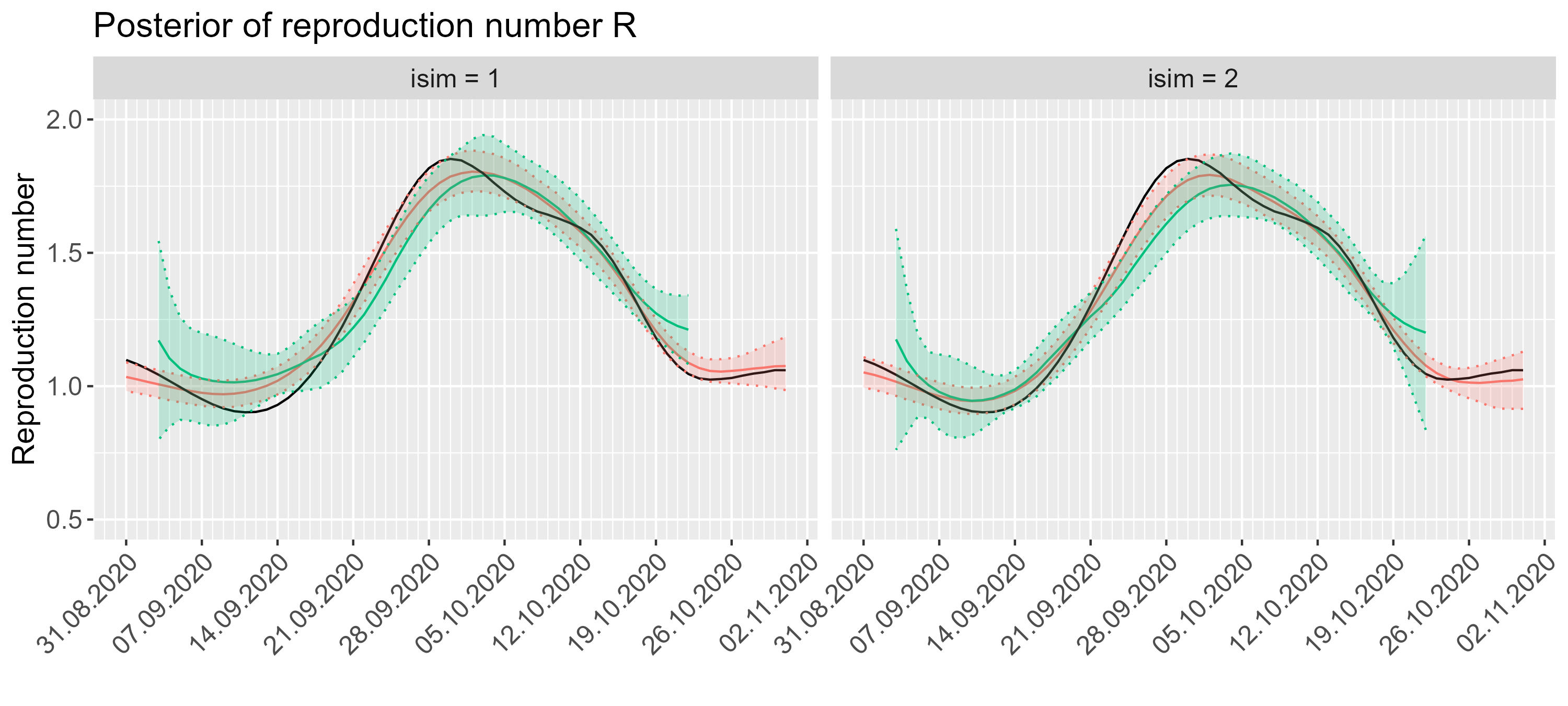}
    \includegraphics[width=\textwidth]{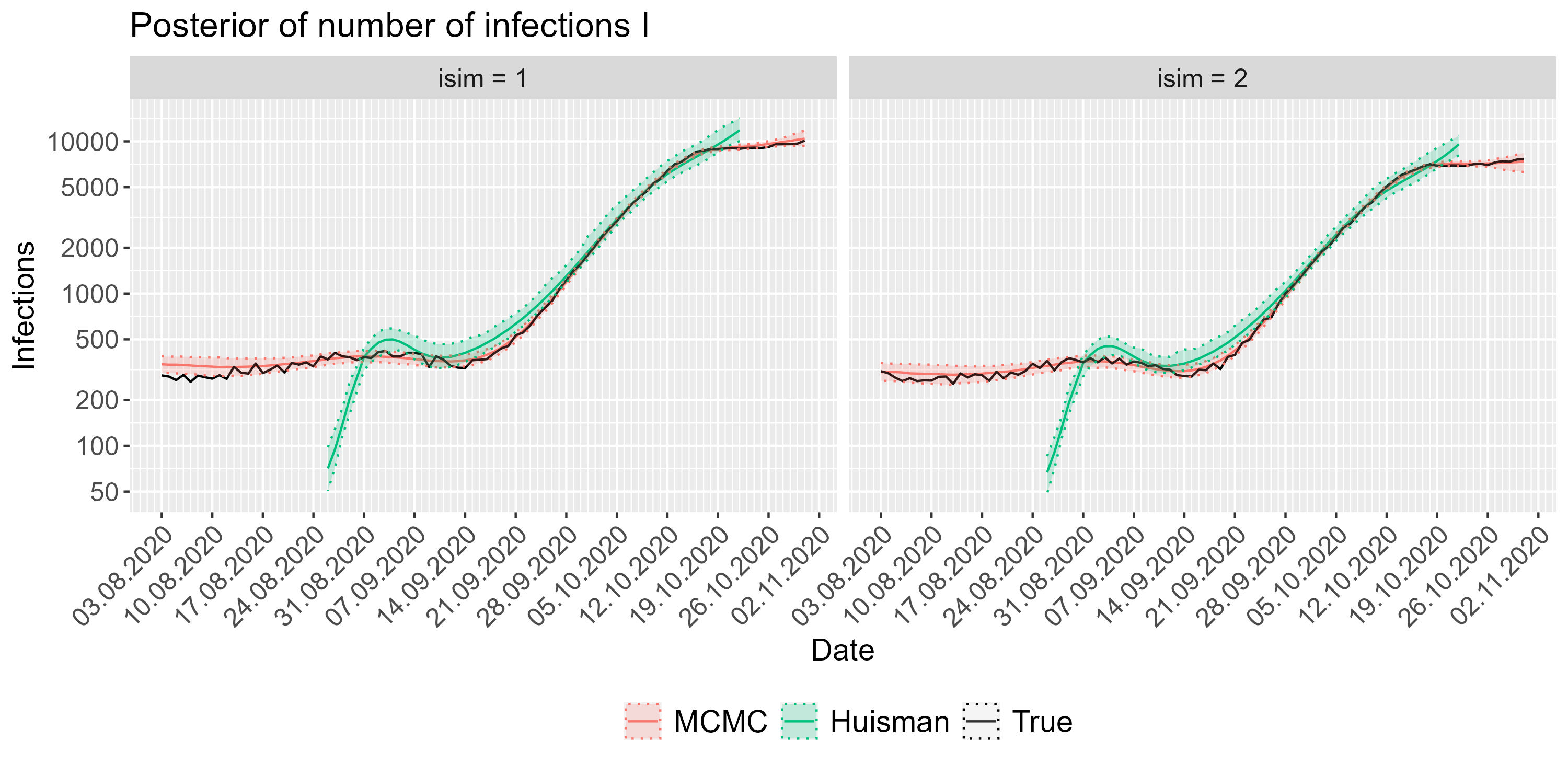}
    \caption{Posterior 2.5\%, 50\% and 97.5\% quantiles of reproduction numbers and incidences given $D_{1:63}$ estimated by our method and that of \citet{huisman2022estimation} for the first two simulated datasets.}  
		\label{post_sim_time}
   \end{center}
\end{figure}

Fig. \ref{post_sim_time} shows posterior quantiles of reproduction and infection numbers estimated using our method and that of \citet{huisman2022estimation} for the first two simulation runs. Figure \ref{post_sim_time_app} in the Supplementary Material \ref{app_simulation} shows the results for additional four simulation runs. We find that point estimates for the number of incidences and the reproduction number $R_e$ obtained with our MCMC sampler are closer to the true values compared to those of \citet{huisman2022estimation}. Furthermore, our intervals are consistently smaller and contain the true values of incidences much more often than those of \citet{huisman2022estimation}. As mentioned in
Section \ref{sub-long}, the approach of \citet{huisman2022estimation} does not provide estimates of $R_e$ and $I$ for some days at the beginning and end of the observation period. But they start to deviate strongly from the correct values even before these self-imposed cut-offs. At the beginning, we see again the effect of zero-padding before preprocessing. 

Next, the accuracy of reproduction number and infection number point and interval estimates for all simulation runs is evaluated using the root mean square error (RMSE) and the interval score given by
\begin{equation*}
  S(l,u;x) = (u-l) + \frac{\alpha}{2} (l-x)\mathbbm{1}_{\{x < l \}} +
  \frac{\alpha}{2} (x-u)\mathbbm{1}_{\{x > u \}},
\end{equation*}
where $l$ and $u$ denote lower and upper endpoint of $95\%$
intervals ($\alpha=0.05$) and $x$ the actual reproduction or infection number. Note that
this scoring function assesses both the coverage frequency which should be as close as possible to $\alpha$ and the width of the 
interval which should be as small as possible. See, e.g.,
\citet{gneiting2007probabilistic} for more information on the interval score.

\begin{figure}[ht!]
  \begin{center}
    \includegraphics[width=\textwidth]{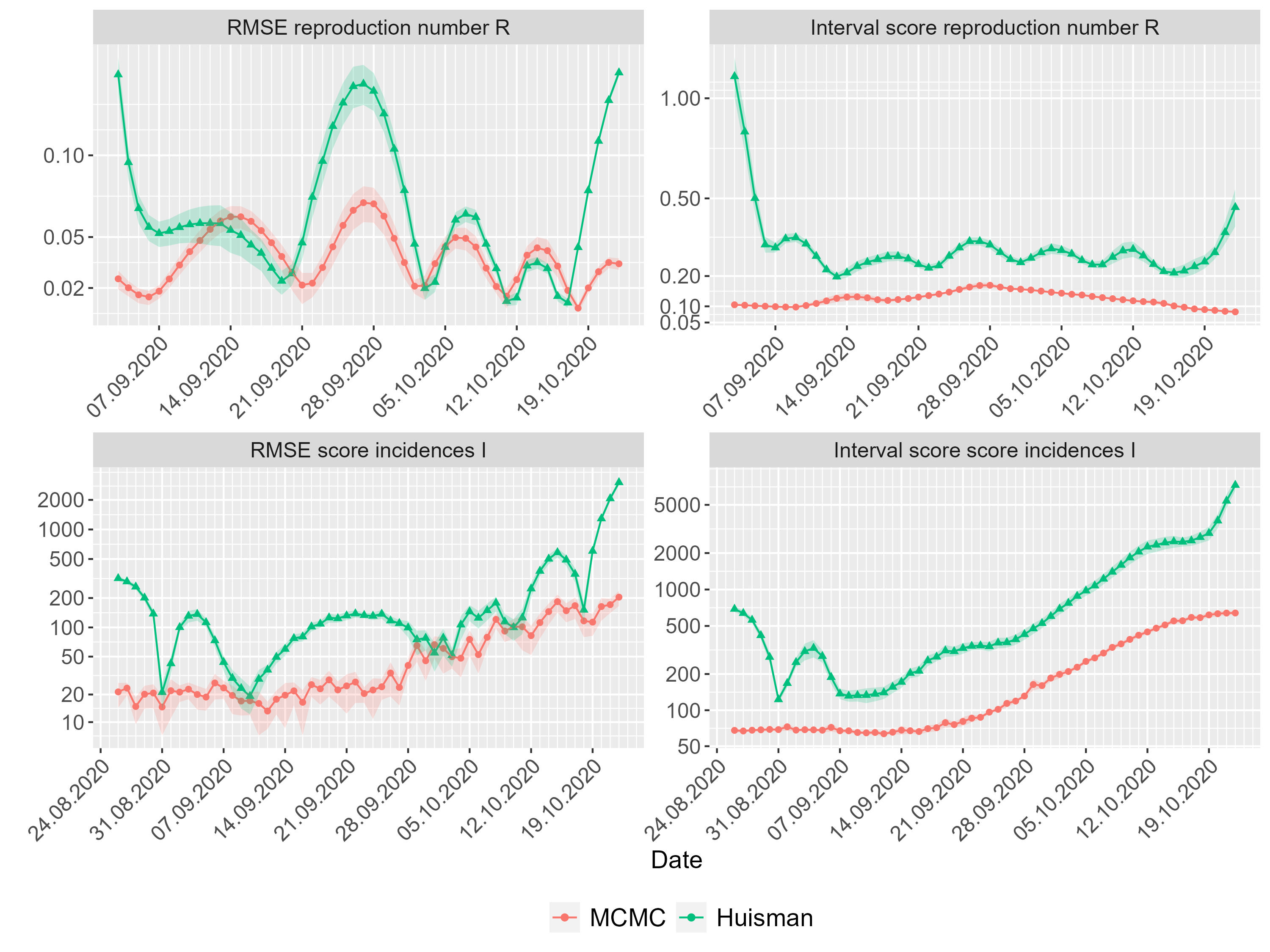}
    \caption{ Comparing the accuracy of our method and that of \citet{huisman2022estimation} in the simulation experiment:
    RMSE and interval score for the reproduction number
      and the number of infections versus time. }  
		\label{results_sim_time}
   \end{center}
\end{figure}

Table \ref{results_sim} reports the results averaged over time and Fig.
\ref{results_sim_time} shows the results per time point, restricted to the period where both methods return estimates. The results confirm that the MCMC algorithm provides both more accurate point estimates and interval estimates compared to the approach of \citet{huisman2022estimation}. Furthermore, the method of \citet{huisman2022estimation} is particularly inaccurate for the most recent days at the end of the observation period.
\input{results_sim.tex}

\begin{figure}[ht!]
  \begin{center}
      \includegraphics[width=\textwidth]{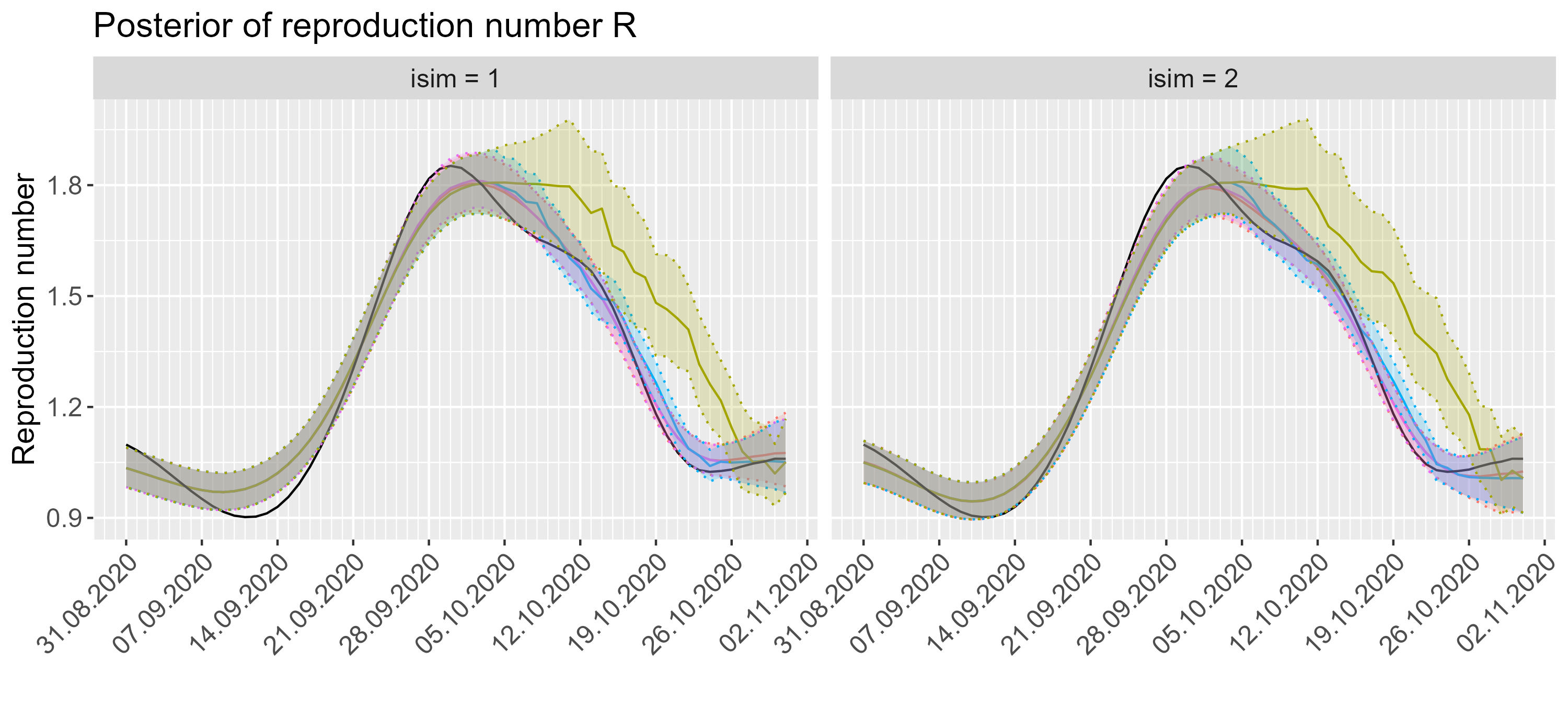}
    \includegraphics[width=\textwidth]{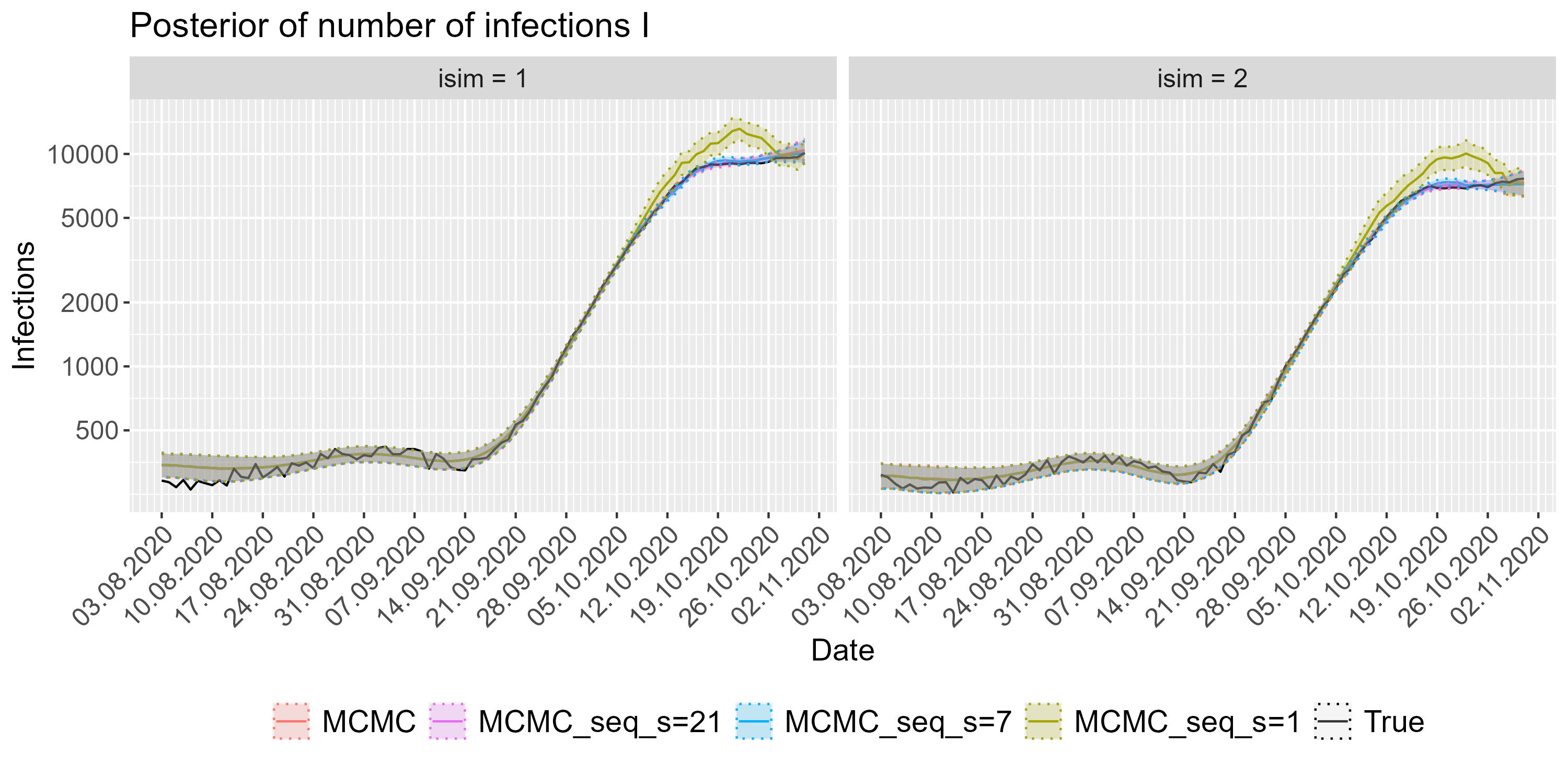}
    \caption{Posterior 2.5\%, 50\% and 97.5\% quantiles of reproduction numbers and incidences for the first two simulated datasets. Quantiles are estimated by MCMC with all 63 observations and sequentially with moving observation windows $D_{(t+s-41):(t+s)}$ for $s=1, 7, 21$.}
 		\label{post_seq}
   \end{center}
 \end{figure}

\begin{figure}[ht!]
  \begin{center}
    \includegraphics[width=\textwidth]{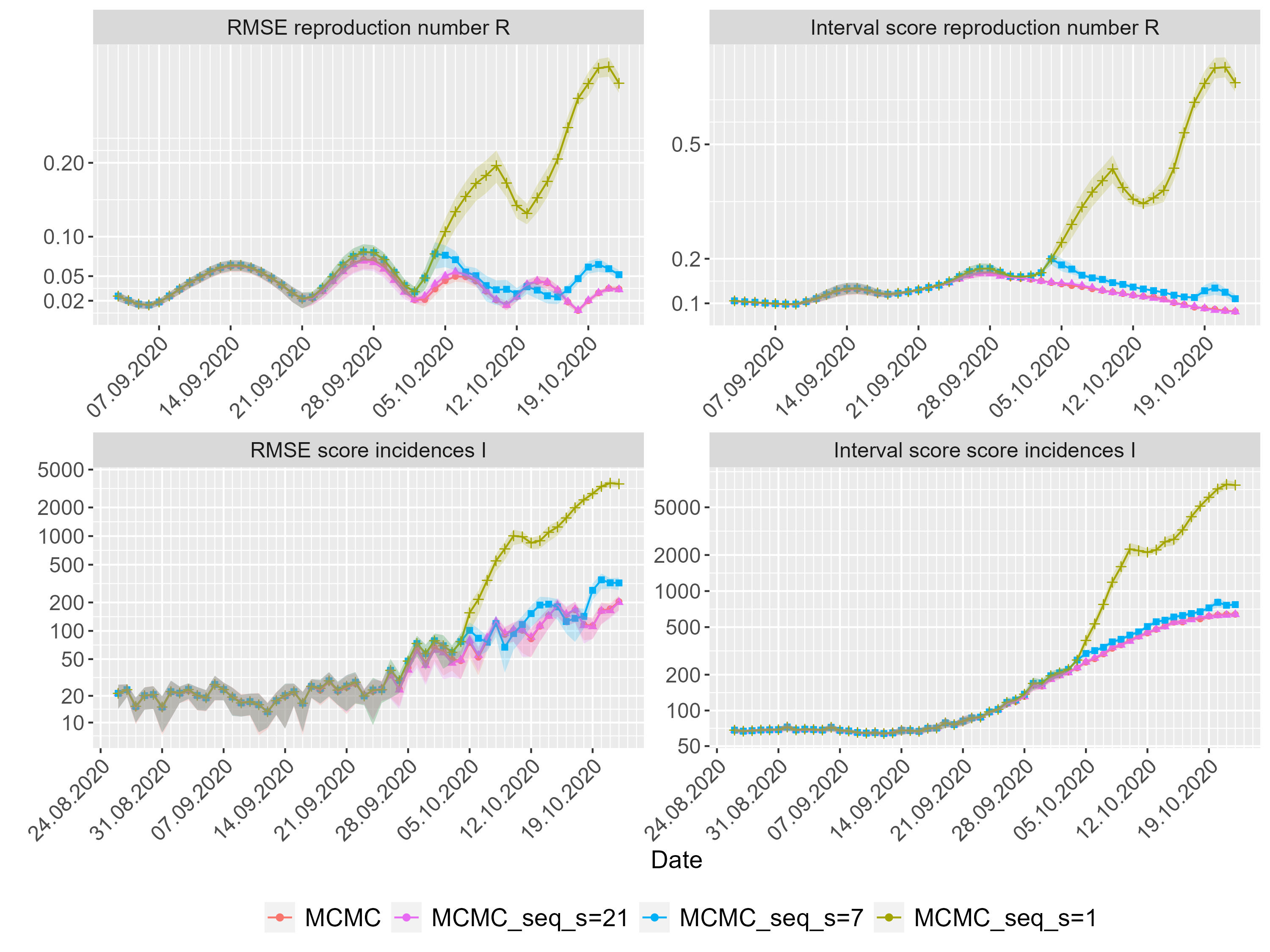}
    \caption{Accuracy measures as in Figure \ref{results_sim_time} for the same methods as in Figure \ref{post_seq}.}
 		\label{results_seq}
   \end{center}
 \end{figure}

Finally, we compare the results of our MCMC algorithm based on all 63 data with the sequential version with a moving window of length 42. Fig. \ref{post_seq} shows estimated quantiles of $R_{e,t}$ and $I_t$ given $D_{1:63}$ and $D_{t+s-41:t+s}$ for $s=1,7,21$ for the same first two simulation runs as above. Fig. \ref{post_seq_app} in Supplementary Material \ref{app_simulation} shows the results for additional four simulation runs. One sees that the sequential estimates of reproduction numbers with $s=1$ decrease more  slowly than the true values in the period 
October 5 to 26. This is presumably due to the chosen prior. This then leads to overestimated incidences about ten days later. But at the very end, the intervals with $s=1$ have been able to catch the truth again. Already with $s=7$ the difference to using all data is small and with $s=21$ there is no visible difference. The accuracy measures of point estimates and intervals over time shown in Fig. \ref{results_seq} confirm these findings for all 100 simulation runs. 

\FloatBarrier
\subsection{Detailed results for data from August 31 until November 8, 2020}
\label{sub-short}

Here we analyze in more detail the performance of sequential
estimation for observations from August 31 (day 1) to November 8, 2020 (day 70). In this period, the number of observed cases changed from an increasing to a decreasing phase. Thus, it is of interest to see how quickly this change becomes visible in the estimates. Moreover, as the government tightened the control measures twice in this period, one would like to know if an effect is visible in our estimates. These tightenings were announced on October 18 (day 49) and on October 28 (day 60) and became effective immediately. They were not as strict as a lockdown, but the size of groups allowed in public places was strongly limited, and the use of masks was compulsory.

In order to separate the effect of changes in the estimated smooth number of cases $\bar{D}_t$ due to the preprocessing from the effect of changes in the spread of the epidemic, we apply the sequential algorithm also to the smooth cases $\bar{D}_t$ estimated from the longer window $1\!:\!84$, assuming that with this longer window the variations due to smoothing are negligible. Note, however, that the latter is not a truly sequential method as the full data was used for preprocessing.

\begin{figure}[ht!]
  \begin{center}
    \includegraphics[width=0.7\textwidth, height=8cm]{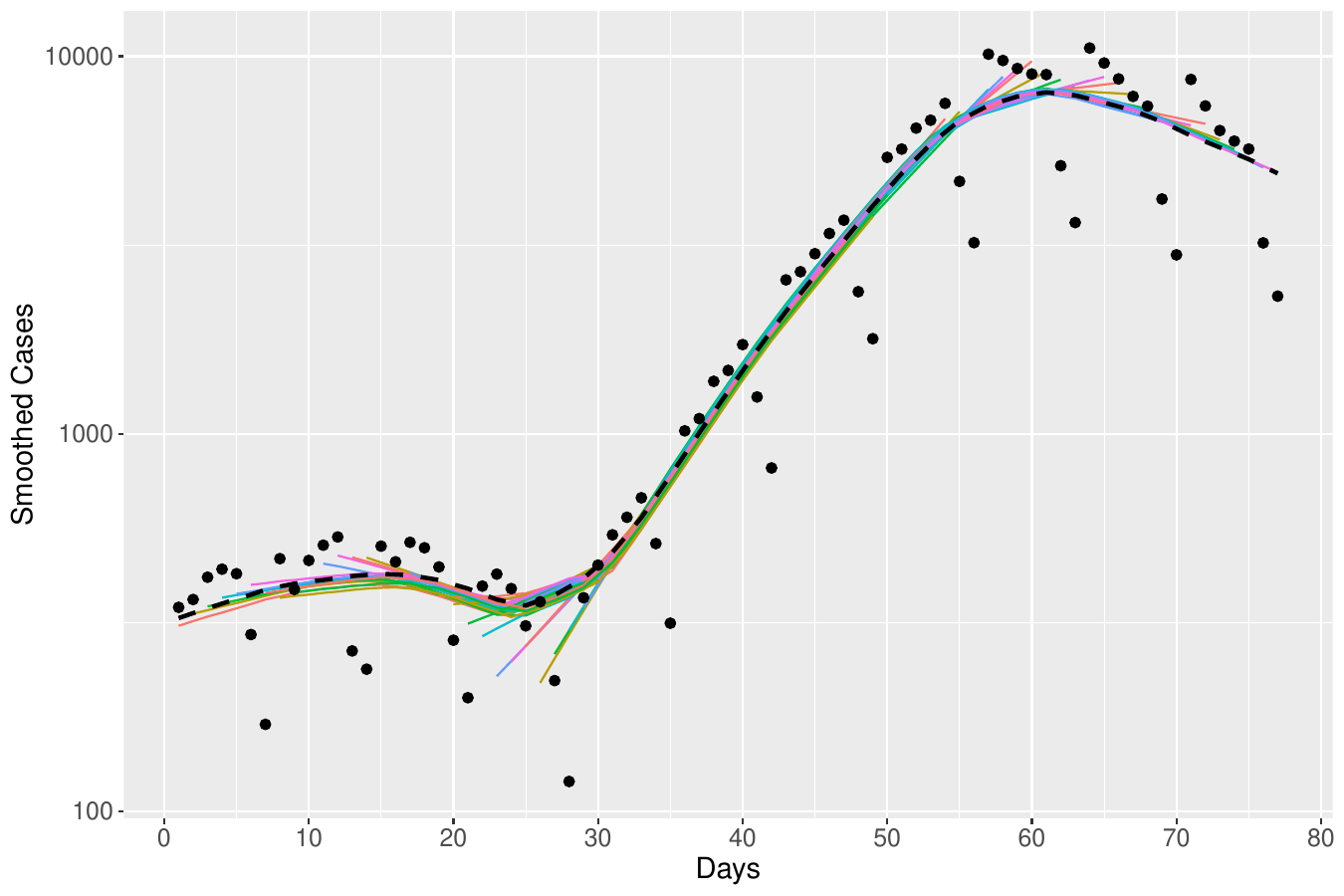}
    \caption{Comparison of sequential and full-data preprocessing by the function \texttt{stl} in $R$ with t.window=15 and 
    s.window= ''periodic''.  Black dots: Raw number of cases.
      Dashed bold black: smoothed cases based on $D_{1:84}$.
      Colored lines: smoothed cases based on $D_{(t-41):t}$ for
      $t=42$ (orange), $t=43$ (ochre), $t=44$ (green), $t=45$ (cyan), $t=46$ (blue), $t=47$ (magenta), then cyclically until $t=77$.}
 		\label{smoothed-cases}
   \end{center}
 \end{figure}

We first compare the sequential and full-data preprocessing methods. As we want to compare predicted values of $\bar{D}_t$ up to seven days ahead, we preprocess the observations in 36 short moving windows from $D_{1:42}$ up to $D_{36:77}$. Fig. \ref{smoothed-cases} shows how much the estimated smooth versions differ among these 36 short and the single long window. If observations are smoothed on the full window $1\!:\!84$, there is a maximum on day 61. In order to observe this also when performing sequential preprocessing with data from a short window, we need observations at least up to day 68. Similarly, to see the local minimum on day 25, one needs observations to start on day 18 or earlier. Thus the uncertainty from smoothing the raw data can persist at both ends for six or more days. 
The estimated weekday effects based on different data windows are shown in Fig. \ref{weekdays} of the Supplementary Material \ref{app-weekday}.

Figs. \ref{r-number-15-77} and \ref{infections-15-77} 
 show estimated posterior quantiles of $I_s$ and  $R_{e,s}$, respectively, against day $s \in 14\!:\!76$. Colors are used to indicate different conditioning windows: For fixed $s$, we condition on $D_{(t-41):t}$ for $\max(42, s-6) \leq  t \leq \min(70, s+28)$ . That is, we look four weeks back and one week ahead from the last observation in each window. Smoothing of the observations was done sequentially using only the data in the same window. We find that a new observation can affect the posterior quantiles up to three weeks earlier. Similarly to the results seen in Section \ref{simulation}, the 95\%-intervals for $R_{e,s}$ in the period $30\!:\!60$ move steadily downward as the conditioning window $(t-41)\!:\!t$ moves further to the right.  This leads then to the same behavior of the incidences about ten days later. The change from a growing to a decreasing phase thus started some time before the interventions became
effective. However, for several days after the first intervention another rise in the reproduction number cannot be excluded. The interventions could have contributed to prevent this from happening. 

\begin{figure}[ht!]
  \begin{center}
    \includegraphics[width=0.8\textwidth]{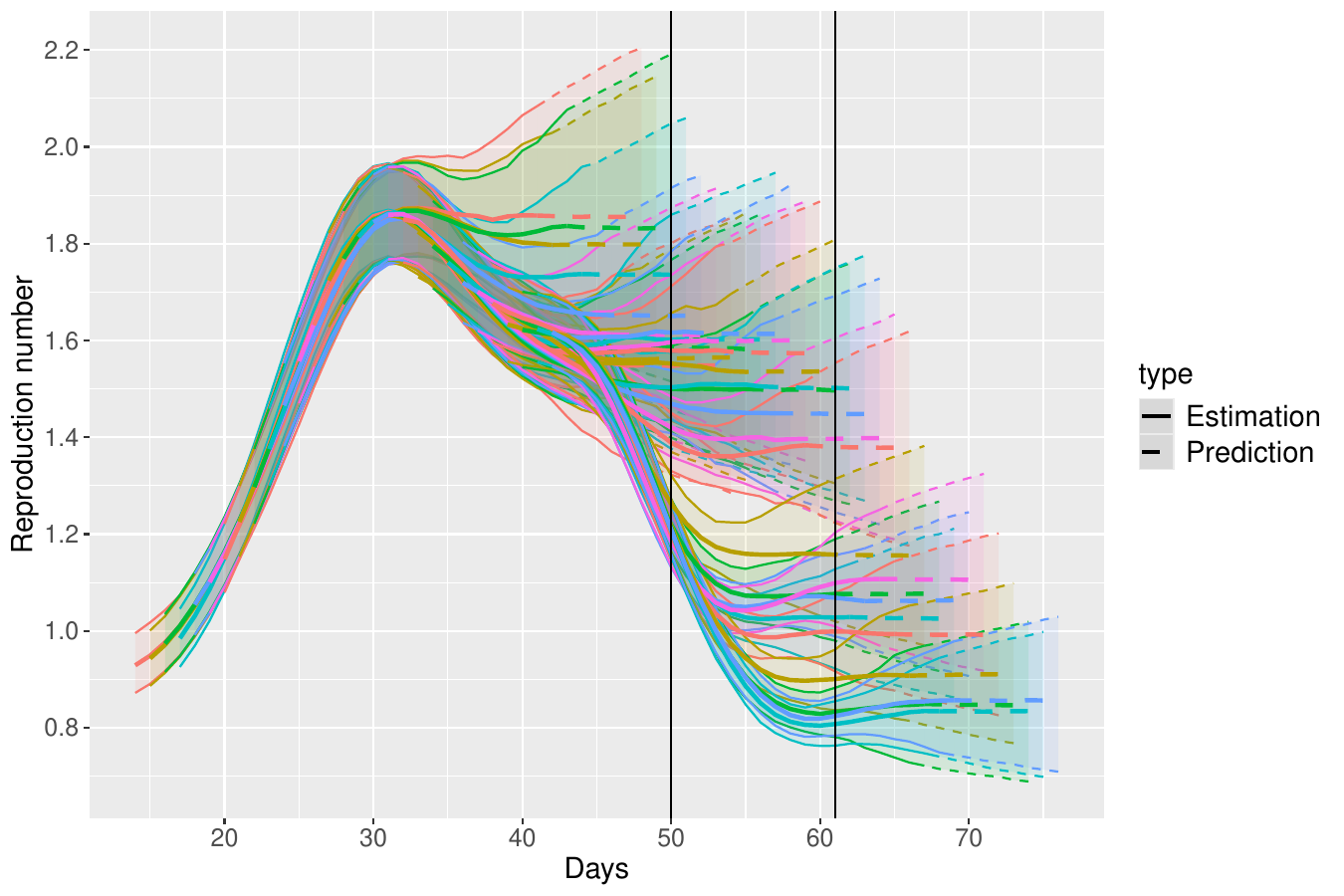}
    \caption{Estimated 2.5\%-, 50\%- and 97.5\% posterior quantiles of reproduction numbers on days $s$ given smooth data $\bar {D}_{(t-41):t}$ for $t \in \max(42, s-6)\!:\!\min(70,s+28)$, distinguished by colors as in Figure \ref{smoothed-cases}. Dashed lines are used for predictive estimates, i.e. for $s \geq t$. Smoothed data are computed from raw data in the same window. The vertical lines mark the days where control measures were tightened.}
   
    \label{r-number-15-77}
   \end{center}
 \end{figure}

\begin{figure}[ht!]
  \begin{center}
    \includegraphics[width=0.8\textwidth]
    {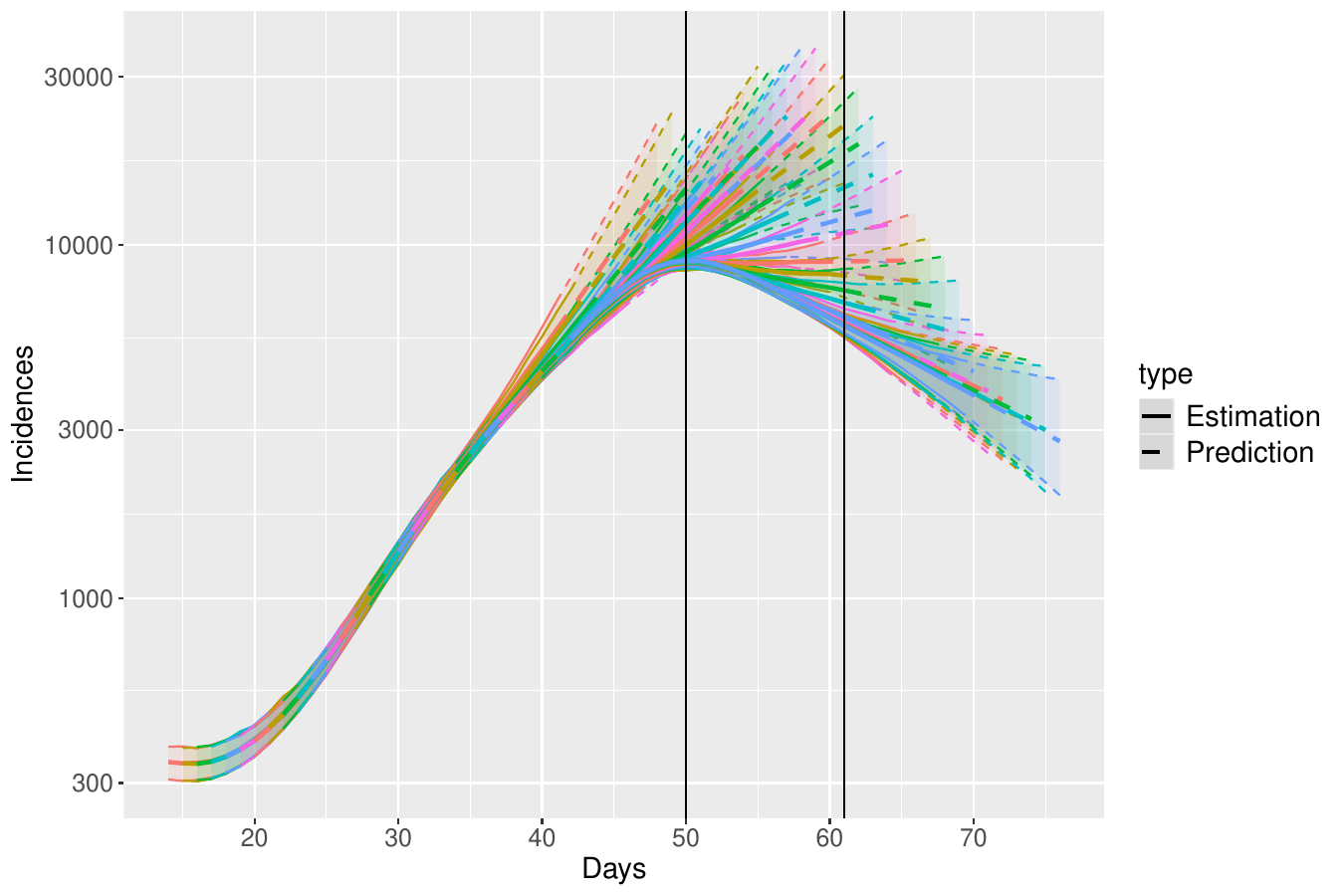}
    \caption{Same as Figure \ref{r-number-15-77} for infections.}
    		\label{infections-15-77}
   \end{center}
 \end{figure}

 Figs. \ref{infections-15-77-a} and \ref{r-number-15-77-a} in the Supplementary Material \ref{app_sub_short} show the same results when observations are smoothed in the full window $1\!:\!84$. As large differences remain between the intervals on the same day based on different data windows, little can be gained by using more sophisticated smoothing methods.

\begin{figure}[ht!]
  \begin{center}
      \includegraphics[width=0.48\textwidth]{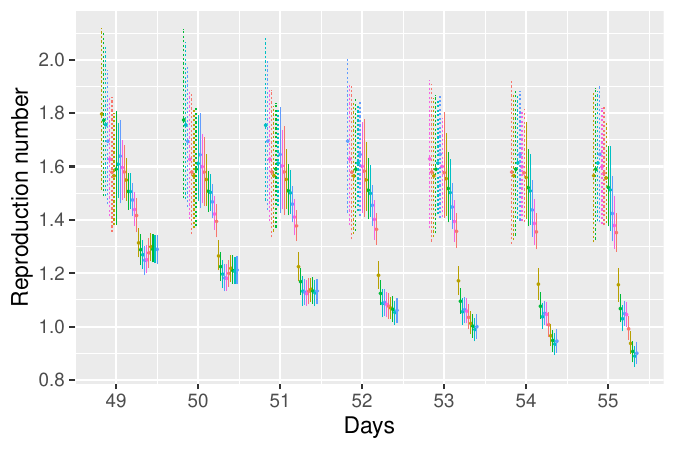}
    \includegraphics[width=0.48\textwidth]{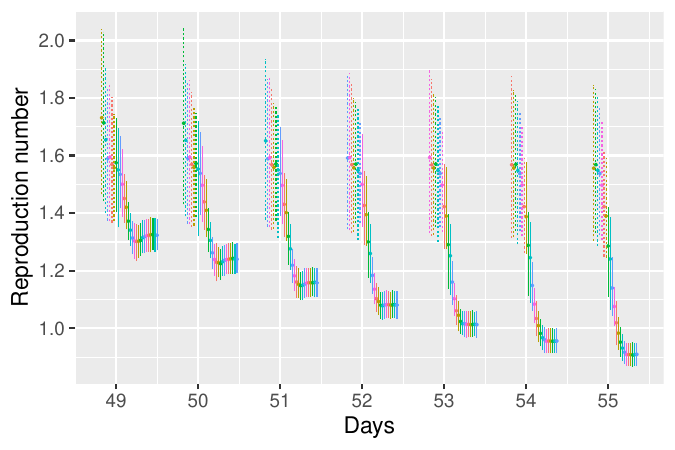}
    \includegraphics[width=0.48\textwidth]{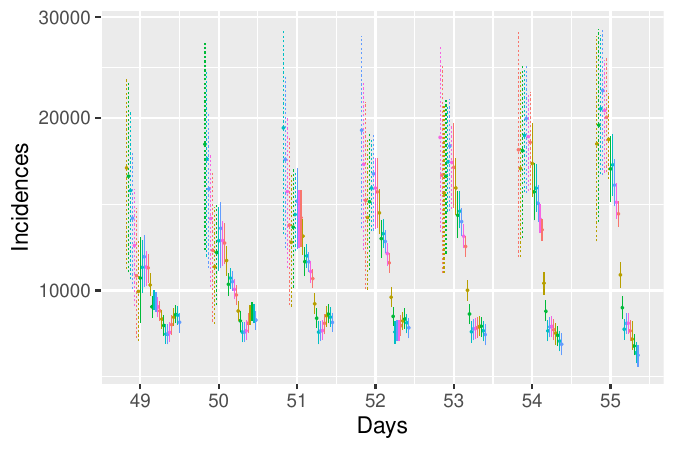}
    \includegraphics[width=0.48\textwidth]{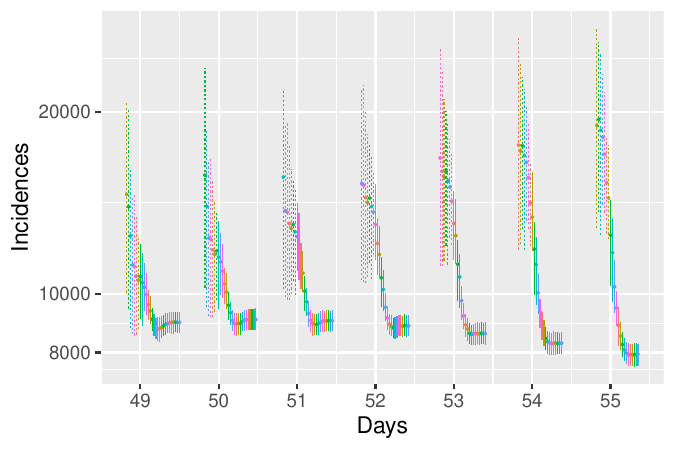}
    \caption{Estimated 2.5\%-, 50\%- and 97.5\% posterior quantiles of reproduction numbers and infections on days $s \in 49\!:\!55$ given smooth data $\bar{D}_{(t-41):t}$ for $t \in (s-6)\!:\!70$, distinguished
      by small horizontal shifts and colors as in
      Fig. \ref{r-number-15-77}. Dashed lines are used for predictive estimates, i.e. for $s \geq t$. Left: Smooth data computed from raw data in the same window. Right: Smooth data computed from raw data in window $1\!:\!84$.}
    		\label{infections-49-55}
   \end{center}
 \end{figure}

 In these figures, it is difficult to see how the estimated quantiles of a variable on a fixed day change as the available observations move forward.  In Fig. \ref{infections-49-55} we slightly shift these quantiles horizontally for different data windows and  for the two different ways of smoothing the observations. As this requires more space, only the results on days 49--56 are shown. When the conditioning window moves forward, the 95\%-intervals typically move downward and become narrower. These changes occur most quickly when the observations on days 57--63 become available. They are larger if we use the same data window for smoothing and estimation. The same plots for the two weeks 42--48 and 56--62 are shown in the Supplementary Material \ref{app_sub_short}.

One can criticize that the posterior intervals for the same day based on different windows often do not overlap. This is, however, not surprising since the influence of the prior for $L$ does not disappear and since the model does not take many important features
into account, e.g. control policies, undetected infections or large differences in the behavior of individuals. 

 \begin{figure}[ht!]
  \begin{center}
    \includegraphics[width=0.48\textwidth]{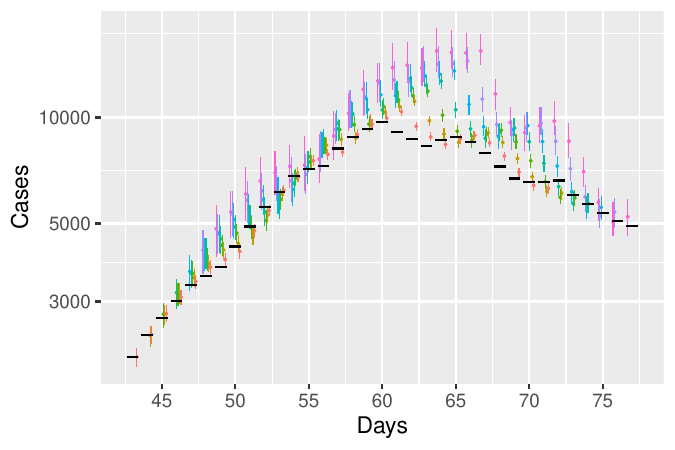}
    \includegraphics[width=0.48\textwidth]{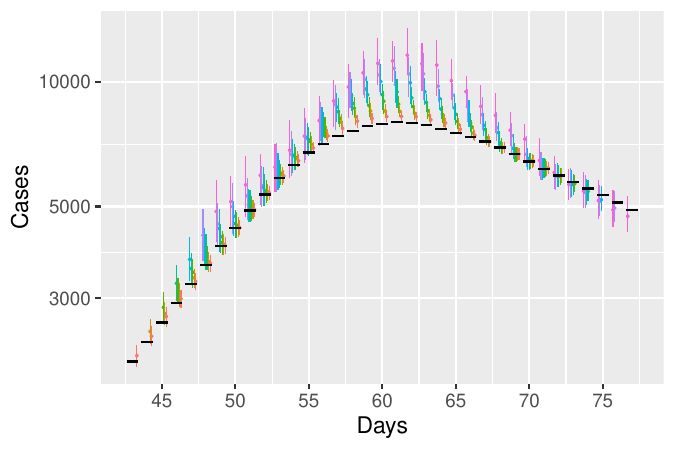}
    \caption{2.5\%-, 50\%- and 97.5\%- prediction
      quantiles of $\bar{D}_t$  given $\bar{D}_{(t-k-41):(t-k)}$ for
      $k \in 1\!:\!7$, shifted horizontally (orange: $k=1$,
      ocher: $k=2$, green: $k=3$, cyan: $k=4$, blue: $k=5$, violet: $k=6$,
      magenta: $k=7$), and true values $\bar{D}_t$ (black horizontal segments) against $t \in 43\!:\!77$. Left: Smooth data computed from raw data in the same window. Right: Smooth data computed from raw data in window $1\!:\!84$. }
    \label{predict-d-43-77}
   \end{center}
 \end{figure}

 Finally, Fig. \ref{predict-d-43-77} compares the predicted
quantiles for $\bar{D}_t$ based on the windows $(t-k-41)\!:\!(t-k)$ for $k=1\!:\!7$, again with the two alternatives to compute the values $\bar{D}_t$. Predictions for more than one step ahead are often far from the truth, even when we use consistently smoothed observations based on the long window $1\!:\!84$. One-step ahead predictions are more reliable except on days 56 and 61--69 when the raw data in the same window are smoothed and on days 56--59 when the raw data on the long window are smoothed. This discrepancy is also the main reason why we had difficulties in developing Sequential Monte Carlo methods, see Section \ref{sub-seq}, as it leads to unbalanced weights in the weighting step. It can again be attributed at least partially to the effect of the additional government restrictions, which are not taken into account by the model used for prediction.

 \section{Summary and discussion}
 \label{sec-disc}
 
We have developed an MCMC algorithm that can produce joint posterior samples for daily reproduction numbers and new infections in a conceptually simple self-exciting model for an epidemic. It provides point estimates and credible intervals that are consistent with the underlying model and a smoothness prior for the daily reproduction numbers. Developing an MCMC algorithm that converges in a reasonably fast manner was challenging. For the Covid-19 data of Switzerland, there are some discrepancies between the results obtained by our method and that
of \citet{huisman2022estimation}. When applied to data simulated according to the
underlying model given a realistic time series of reproduction
numbers, point and interval estimates obtained by our method are more
accurate than those by the approach of \citet{huisman2022estimation}.

In order to have an efficient algorithm when new data become available each day, we use a moving window of the most recent six weeks of data. As a new observation has practically no effect on estimates on days that are more than three weeks in the past, we obtain smooth trajectories of posterior quantiles over the whole past by updating the most recent three weeks.

An additional difficulty arises with real data because they show strong
weekday and holiday effects. We have chosen to remove these effects and
to smooth the data before running our MCMC algorithm. In future work, one might try to incorporate this step also in the MCMC algorithm, but one would have to find plausible assumptions on a time-varying delay distribution which models also holiday effects and slow changes in weekday effects.

In our examples, the credible intervals near the end of a period typically change substantially as new observations become available. Often they do not even overlap. This might be due to  model deficiencies and/or the inadequacy of the chosen prior for the reproduction numbers. Future research can investigate how to obtain more stable credible intervals, for instance with a data-dependent choice of the hyperparameter $\tau$ such as cross-validation or with a more complex prior. Future research can also analyze how various forms of misspecification affect the model and the estimates considered in this article.

Our MCMC algorithm can be adapted to many different extensions of the underlying model. For instance, when the number of detected cases are stratified by age or geographical units, we could also stratify the
incidences and reproduction numbers in the same way. But then we would
have to make assumptions how the different strata interact in the dynamics. A different extension could try to estimate also a time-varying probability of undetected infections, using the observed percentage of negative test results or data on the abundance of the virus in sewage plants. Such extensions are beyond the scope of this
article. In any case, comparing the results based on more complex and detailed models with those based on simpler assumptions as ours is of interest and can provide additional insights.

%% file: results_sim.tex
\begin{table}[ht]
\centering
\begin{tabular}{rll}
  \hline
 & Posterior MCMC & Huisman \\ 
  \hline
RMSE R & 0.038 & 0.0663 \\ 
  Interval score R & 0.1235 & 0.306 \\ 
  RMSE I & 57.1 & 255 \\ 
  Interval score I & 212.2 & 1041.9 \\ 
   \hline
\end{tabular}
\caption{Accuracy of point estimates (RMSE) and interval estimates (interval score) for the reproduction number (R) 
             and the number of infections (I) averaged over time.} 
\label{results_sim}
\end{table}

%% file: appendix.tex
\begin{appendix}
\setcounter{figure}{0}
\renewcommand{\thefigure}{A\arabic{figure}}
\setcounter{equation}{0}
\renewcommand{\theequation}{A\arabic{equation}}
\section*{Supplementary Material}	
\section{Proof of Lemma 1}
\label{app-lemma1}

By \eqref{eq:model-I} $E[I_t]$ satisfies the following linear recursion
\begin{equation}
  \label{eq:recursion}
  \psi_t = R_e \sum_{k=1}^{K_w} w_k \psi_{t-k},  \quad t=K_w+1, K_w+2, .... .
\end{equation}
The set of all solutions for this recursion is a linear space of dimension $K_w$. A sequence of the form $\psi_t=\lambda^t$ is a solution iff 
$$ P(\lambda) = \lambda^{K_w} - R_e\sum_{k=1}^{K_w} w_k \lambda^{K_w-k}=0.$$
If $\lambda$ is an $\ell$-fold zero of $P(\lambda)$, then 
$\psi_t = t^j \lambda^t$ satisfies \eqref{eq:recursion} for $0 \leq j \leq \ell$. Hence, we always have $K_w$ particular solutions that we denote by $\psi^{(j)}_t$ \citep[][Theorem 2.23, p.77]{elaydi2005introduction}.
Because they increase at different rates for $t \rightarrow \infty$, they are also linearly independent and thus form a basis.

Since $P(\lambda) \lambda^{-K_w}$ increases strictly monotonically
 for $\lambda >0$ with limits $-\infty$ and $1$ as $\lambda$
tends to $ 0$ and $\infty$, respectively, there is a unique positive zero of $P(\lambda)$ that we call $\rho$. Moreover, for any other zero $\lambda$ it holds that
$$1 = R_e\left|\sum_{k=1}^{K_w} w_k \lambda^{-k}\right| < 
R_e\sum_{k=1}^{K_w} w_k |\lambda|^{-k},$$
since by assumption all $w_k >0$ (having two consecutive positive
$w_k$'s is sufficient). 
This implies $P(|\lambda|) < 0$ and thus $|\lambda|<\rho$. We set
$\psi^{(1)}_t = \rho^t$.

In order to complete the proof of the first part, we have to show
that the coefficient $u_1$ is positive for any solution $(\psi_t)$ whose $K_w$ initial conditions are all non-negative and not identically zero. This holds because the recursion implies that 
$\psi_t > 0$ for $t>K_w$ and thus $\psi_t > c \rho^t$ for some 
$c>0$ and $K_w < t \leq 2K_w$. The recursion implies then that
$\psi_t > c \rho^t$ for all $t>K_w$. 

For the second part, we write the recursion \eqref{eq:recursion} in matrix form as
$$\boldsymbol{\psi}_{t+1} = A \boldsymbol{\psi}_t, \quad t=K_w, K_w+ 1, \ldots,$$
where $\boldsymbol{\psi}_t = (\psi_t, \ldots, \psi_{t+1-K_w})^T$ and $A_{1k} = w_k$, 
$A_{ik}=\delta(i-1,k)$ for $i>1$.

It follows from the law of iterated expectations
that $C_t = \Cov(\mb I_t)$ satisfies the recursion
$$C_t = E[\Cov(\mb I_t \mid \mb I_{t-1})] + \Cov(E[\mb I_t \mid \mb I_{t-1}]) = E[I_t] \mb e \mb e^T + \Cov(A \mb I_{t-1}) = E[I_t] \mb e \mb e^T + A C_{t-1}A^T$$
for $t>K_w$ where $\mb e = (1, 0, \ldots 0)^T$. Iterating this recursion leads to
$$ C_t = \sum_{s=K_w+1}^t E[I_s] (A^{t-s} \mb e)(A^{t-s} \mb e)^T
+ A^{t-K_w} \mb C_{K_w} (A^{t-K_w})^T \quad t=K_w +1, \ldots.$$
By the arguments used for the first part, 
$$A^{t}\mb e = \sum_j u_j A^{t} \mb \psi^{(j)}_{K_w}
= \sum_{j=1}^{K_w} u_j \mb \psi^{(j)}_{K_w + t}$$
with unique coefficients $u_j$ and $u_1 > 0$. Similarly
$$A^{t-K_w} \mb I_{K_w} = \sum u_j \mb \psi^{(j)}_t$$
where the random coefficients $u_j$ are obtained by a linear
transformation of the initial values $\mb I_{K_w}$. From this it follows
that $\rho^{-2t} C_t $ converges to a positive constant times
$\mb \psi^{(1)}_{K_w}(\mb \psi^{(1)}_{K_w})^T$.
This finishes the proof of the second part.

\section{Values of the infectivity profile and the delay distribution}
\label{app-delay-comp}
The infectivity profile we chose is 
$$w_k = \frac{F(k + 0.5) - F(k -0.5)}{1 - F(12.5)} \quad (1 < k \leq 12), \quad w_1=\frac{F(1.5)}{1 - F(12.5)}$$
where $F$ is the cumulative distribution function of a Gamma random
variable with shape 
$\alpha=(4.8/2.3)^2$ and rate $\lambda=4.8/2.3^2
$

The delay between infection and detection is chosen to be the sum of two independent Gamma random variables with shape parameters
$\alpha_1=(5.3/3.2)^2$, $\alpha_2=(5.5/3.8)^2$ and rate parameters
$\lambda_1=5.3/2.2^2$, $\lambda_2=5.5/3.8^2$, discretized to
integers from 1 to 28. Thus we set
$$m_k = \frac{G(k + 0.5) - G(k -0.5)}{1 - G(28.5)} \quad (1 < k \leq 28), \quad m_1=\frac{G(1.5)}{1 - G(28.5)}.$$
Here 
$$G(x) = \int_0^x F(x - y;\alpha_2, \lambda_2) f(y; \alpha_1, \lambda_1)dy $$
where $f(x;\alpha,\lambda)$ and $F(x; \alpha, \lambda)$ are the density
and distribution function for the Gamma distribution with shape 
$\alpha$ and rate $\lambda$. We compute $G$ using numerical integration. 

The following table shows the values of the infectivity profile and the delay distribution:

\begin{center}
\begin{tabular}{c|rrrrrrrrrrrrrr}
\
$k$ &  1 & 2 & 3 & 4 & 5 & 6 & 7 & 8 & 9 & 10 & 11 & 12 & 13 & 14 \\
$1000 \cdot m_k$ & 1 & 7 & 20 & 38 & 57 & 73 & 83 & 88 & 89 & 85 & 78 & 69 & 60 & 51\\ 
$1000 \cdot w_k$ & 31 & 114 & 179 & 190 & 163 & 122 &  83 & 53 & 32 & 18 & 10  & 5  & - & - \\ 
\hline
$k$ & 15 & 16 & 17 & 18 & 19 & 20 & 21 & 22 & 23 & 24 & 25 & 26 & 27 & 28 \\
$1000 \cdot m_k$ & 43 & 35 & 28 & 23 & 18 & 14 & 11 & 8 & 6 & 5 & 4 & 3 & 2 & 1\\ 
\end{tabular}
\end{center}

\section{Details about the MCMC algorithm}

\subsection{Derivation of the acceptance ratio
in Subsection \ref{sub-sample-I}}
\label{app-acc-ratio}
By definition of $\nu^*_{s,t}$ we have
$\log \nu^*_{s,t} = \log \psi_s^* - \log \pi_t^* + \log m_{s,t}$.
Therefore it follows
from $\sum_{s= t-K_m}^{t-1}A^*_{s,t} = D_t$ and 
$\sum_{t=1}^T A^*_{s,t} = B^*_s$ that
\begin{equation}
\label{eq:acc-ratio-1}
     \sum_{t=1}^T \sum_{s=t-K_m}^{t-1} A^*_{s,t} \log \nu^*_{s,t}
 = \sum_{s=1-K_m}^{T-1} B^*_s \log \psi^*_s + 
 \sum_{t=1}^T \sum_{s=t-K_m}^{t-1} A^*_{s,t} \log m_{s,t} -
      \sum_{t=1}^T D_t \log \pi^*_t
\end{equation}

Using $U^*_s=I^*_s - B^*_s$ we then obtain
\begin{eqnarray*}
    \log q(\mb I^*,\mb A^*, \mb I^*_{init}\mid \mb L, \mb D, \mb I)
  &=&  \sum_{s=1-K_m-K_w}^{-K_m} 
      (I^*_s \log \lambda_s -\lambda_s - \log I^*_s!)\\
  &+& \sum_{s=1-K_m}^{T-1} (I^*_s \log \lambda^*_s -
      B^*_s\log(\lambda^*_s/\psi^*_s) +  U^*_s \log u_s - u_s\lambda^*_s
      - \log U^*_s!)\\
  &+& \sum_{t=1}^T \sum_{s=t-K_m}^{t-1} (A^*_{s,t} \log m_{s,t} -
      \log A_{s,t}^*!) + \sum_{t=1}^T (\log D_t! - D_t \log \pi_t^*).
\end{eqnarray*}
Combining this with
$$\log p(\mb I^*, \mb I^*_{init}\mid \mb L) = \sum_{s=1-K_m-K_w}^{-K_m} 
(I^*_s \log \lambda_s -\lambda_s - \log I^*_s!) + 
\sum_{s=1-K_m}^{T-1} (I^*_s \log \lambda_s^* -\lambda^*_s - \log I^*_s!)$$
and
\begin{eqnarray*}
  \log p(\mb A^*\mid \mb I^*)
  &=& \sum_{s=1-K.m}^{T-1}(\log I^*_s! + U^*_s\log u_s - \log U^*_s!)\\
  &+& \sum_{t=1}^T \sum_{s=t-K_m}^{t-1} (A^*_{s,t} \log m_{s,t} -
      \log A_{s,t}^*!)
\end{eqnarray*}
we obtain
$$ \log\left(\frac{p(\mb I^*, I^*_{init}  \mid \mb L) 
   p(\mb A^*\mid \mb I^*)}
  {q(\mb I^*,\mb A^*, \mb I^*_{init} \mid \mb L, \mb D, \mb I)}\right)
  = \sum_{s=1-K_m}^{T-1} (B^*_s \log(\lambda^*_s/\psi^*_s) -
  b_s\lambda^*_s) + \sum_{t=1}^T (-\log D_t! + D_t \log \pi_t^*).
  $$
  From this \eqref{eq:prop1-r} follows. 
  
  There is an alternative formula 
for the acceptance ratio \eqref{eq:prop1-r} involving 
the Poisson log likelihood ratio
$$\ell_{Pois}(x; \lambda, \psi) = \log f_{Pois}(x,\lambda) -
\log f_{Pois}(x,\psi) = x \log (\lambda/\psi) - (\lambda - \psi).$$
Using 
\begin{equation}
\label{eq:acc-ratio-2}
    \sum_t \pi^*_t = \sum_s\psi^*_s \sum_t m_{s,t}=\sum_s \psi^*_s b_s
\end{equation}   
it follows  that
\begin{eqnarray}
  \label{eq:prop1-lr}
  \log r(\mb I^*, \mb I, \mb A^*, \mb A, \mb L, \mb D)
  &=&  \sum_{s=1-K_m}^{T-1} (\ell_{Pois}(B^*_s;b_s\lambda^*_s,b_s\psi^*_s) +
      \ell_{Pois}(B_s;b_s\psi_s,b_s\lambda_s)) \nonumber \\
       & & + \sum_{t=1}^T \ell_{Pois}(D_t;\pi^*_t,\pi_t).
\end{eqnarray}
This has the following interpretation: Under the target, $B_s^*$ has
approximately a Poisson($b_s\lambda^*_s)$-distribution, whereas under
the proposal it has approximately a Poisson($b_s \psi^*_s)$-distribution.
Thus the first term is close to zero if $B_s^*$ also fits to the target
and the second term is close to zero if $B_s$ also fits to the proposal.
Finally, the third term is positive if 
the fit of $\pi^*_t$ for $D_t$ is better than the fit of $\pi_t$.

\subsection{Prior means of infections for the first $K_w$ days}
\label{app-prior-mean-I}

When choosing the prior means $\lambda^0_s$ of $I_s$ for
$1-K_m - K_w <s \leq -K_m$, care has to be taken that it does not lead to
conflicts with the observations $\mb D$ in order to avoid a very low acceptance
rate of proposals. We assume that $I_t$ evolves according to the
dynamic model \eqref{eq:model-I} from $t=1-K_m$ onwards whereas the
observations start at $t=1$. Hence there is some flexibility for
adjustment if the chosen values $\lambda^0_s$ are slightly off.

When the number of detections are available for $K_m$ days before
the statistical analysis starts, then these values $D_t$ can be used to choose the $\lambda_s$.
This holds in our case, and we take 
$\lambda^0_s$ constant and equal to the mean of $D_{-6}, \ldots D_0$. This
removes the weekday effects and gives a slight bias to plausible values
of $I_s$ near the start of the observations.

In general, one might be forced to use the observations $D_t$ for
$t>0$ to choose the $\lambda^0_s$, although this is strictly speaking not
legitimate because it makes the prior dependent on the data $\mb D$. Then
one can use the same procedure that we use to construct the starting value
$I$ for our MCMC algorithm described next.

\subsection{Initial values of the MCMC algorithm}
\label{app-init-val}
We choose starting values that fit well to the observations $\mb D$. Our
procedure is as follows: For $1 - K_m - K_w \leq s \leq -K_m$ we generate $I_s$ according
to the prior, i.e. $I_s \sim \lambda^0_s$ independently of each other. 
For $-K_m < s < T$, we choose $I_s$ by the following procedure: 
We shift  $\mb D$ back by 10 days, extend it by
constant values on both ends to cover the whole period and 
apply 10 steps of the EM algorithm (see Appendix \ref{app-em}), Finally we smooth the potential jump between $I_{-K.m}$ and $I_{1-K.m}$ and round to the nearest integer. 
 This is very similar to the estimate of $\mb I$ used in
\citet{huisman2022estimation}. The starting value for $\mb L$ is given by
$$L_t = \log \left(\frac{I_t}{\sum_k w_k I_{t-k}}\right).$$
As we need also starting values for $\mb B$ in order to compute the
acceptance ratio \eqref{eq:prop1-r}, we set $B_s=I_sb_s$, rounded to the next integer.

\section{Details about sampling from $p(\mb L \mid \mb I, \mb I_{init})$}
\label{app-prop-Lstar}

Recall that the target is
$$\log p(\mb L \mid \mb I)=- \frac{L_{1-K_m}^2}{2 \sigma^2} -
\sum_{t=1-K_m}^{T-2} \frac{(L_{t+1}-L_t)^2}
{2 \tau^2} + \sum_{t=1-K_m}^{T-1} (I_tL_t - \lambda_t)$$
and -- given the current value $\mb L$ -- we want to propose a new value $\mb L^*$  from a Gaussian distribution that is close to 
$p(\mb L^* \mid \mb I).$
For this, we replace $\lambda^*_t$ by a second-order Taylor approximation
$$\lambda^*_t \approx \lambda_t(L_t^*(1-L_t)+0.5 (L_t^*)^2 + 1 -L_t + 0.5 L_t^2).$$
This leads then to a Gaussian proposal
\begin{equation}
  \label{eq:logq-L}
\log q(\mb L^* \mid \mb I,\mb L) \propto
   -\frac{1}{2}(\mb L^*)^T Q(\mb L) \mb L^*
     + \mb b(\mb L)^T \mb L^*.
\end{equation}
The matrix $Q$ and the vector $\mb b$ are given by
\begin{eqnarray*}
Q(\mb L) &=& Q_L + \diag(\mb \lambda)\\
\mb b(\mb L)&=& \mb I  - (1-\mb L)\mb \lambda
  \end{eqnarray*}
  and $Q_L$ is the precision matrix of the prior for $\mb L$.
The mean of the proposal is $Q(\mb L)^{-1}\mb b(\mb L)$ and the covariance is $ Q(\mb L)^{-1}$. In the following we drop the dependence of $Q$ and $\mb b$ on $\mb L$.

The acceptance probability
$$\min\left(1, \frac{p(\mb L^* \mid \mb I) q(\mb L \mid \mb L^*, \mb I)}
  {p(\mb L \mid \mb I) q(\mb L^* \mid \mb L,\mb I)} \right).$$
depends also on the normalization constant of $q$. 
In order to compute it and draw the new proposal, the Choleski decomposition $Q=U^TU$ is used, where $U$ is an upper triangular matrix. The mean under $q$ is equal to 
$U^{-1}U^{-T}\mb b$, the normalizing constant is 
$(\det Q)^{1/2}=\prod_i U_{ii}$ and for drawing $\mb L^*$ we use
$$\mb Z \sim \N(0,I)  \Rightarrow \mb L^* = U^{-1}(\mb Z + U^{-T}\mb b)
\sim \N(Q^{-1}\mb b,Q^{-1}).$$
Finally, the quadratic form in the exponent of the density is
$$(\mb L^*-Q^{-1} \mb b)^TQ(\mb L^*-Q^{-1}\mb b) = ||U\mb L^* - U^{-T}\mb b||^2 =||\mb Z||^2$$
where $||.||$ is the Euclidean norm ($||\mb Z||^2 = \mb Z^T \mb Z$).

\section{Delay distributions with weekday effects}
\label{app-weekday}

There are at least two ways for constructing such distributions
from a time-invariant delay distribution $m^{(0)}$. The first assumes that
a detection that would normally occur on weekday $i$ occurs $k$ days
later with probability $v(i,k)$ where $k \in 0\!:\!6$. For
given $v$ and $m^{(0)}$ the delay distribution then becomes
$$m_{s,s+k} = \sum_{j=\max(1,k-6)}^k m^{(0)}_j v( (s+j) \bmod 7,k-j).$$

The second way uses a multiplicative modification of $m^{(0)}$. That is,
one increases or decreases $m_{s,t} = m^{(0)}_{t-s}$ depending on the
weekday of $t$ while keeping the sum over $t$ unchanged. For this 
one chooses positive weights $w_{0:6}$ with
$$ \sum_{k=0}^6 w_k = \sum_{k=1}^{K_m} m^{(0)}_k$$
and sets
$$m_{s,s+k}  = w_{(s+k) \bmod 7} \, \frac{m^{(0)}_k}
  {\sum_{k'=k \bmod 7}m^{(0)}_{k'}}.$$
Splitting the sum over $k \in 1\!:\!K_m$ into an outer sum over
$i \in 0\!:\!6$ and an inner sum over $k$ for which $(s + k)=i \bmod 7$
one sees that for any $s$
$$\sum_{k=1}^{K_m} m_{s,s+k} = \sum_{i=0}^6 w(i) = \sum_{k=1}^{K_m} m^{(0)}_k.$$

The second choice has the advantage that if the time series $(I_t)$ has an
exponential trend, then $(D_t)$ has the same trend, up to a multiplicative
constant, and the multiplicative weekday effect $w_{0:6}$. For the proof,
we assume that $E[I_t] = C \exp(\lambda t)$. Then
\begin{eqnarray*}
E[D_t] &=& C \sum_{k=1}^{K_m} m_{t-k,t} \exp(\lambda( t-k)) =
           C \exp(\lambda t) w(t \bmod 7) \sum_{k=1}^{K_m}
           \frac{m^{(0)}_k\exp(-\lambda k)}{\sum_{k'=k \bmod 7} m^{(0)}_{k'}}
           \\
  &=& C' \exp(\lambda t) w(t \bmod 7).
\end{eqnarray*}

\begin{figure}[h!]
  \begin{center}
    \includegraphics[width=0.7\textwidth]{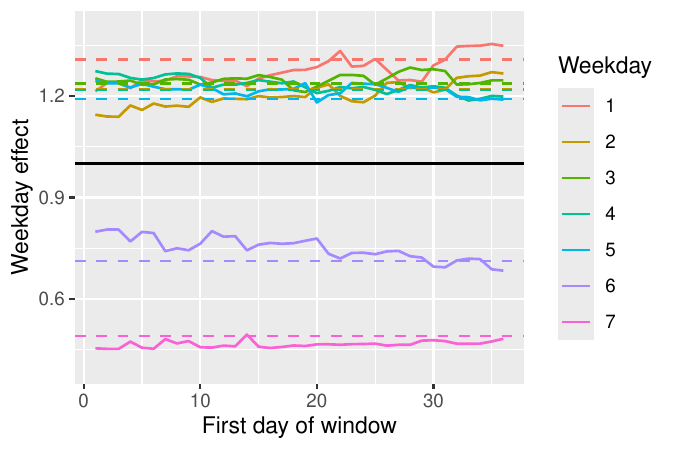}
    \caption{Periodic multiplicative weekday effects estimated from raw data
      $D_{t:(t-41)}$ against $t \in 1\!:\!36$ (solid) and from raw data 
      $D_{1:84}$ (dashed) by the function \texttt{stl} in $R$ with t.window=15 and s.window="periodic". The black line corresponds to  absence of a weekday effect. }
 		\label{weekdays}
   \end{center}
 \end{figure}

The transition matrix $v$ or the weekday distribution $w$ would then have to be estimated together with incidences and reproduction numbers. Figure \ref{weekdays} shows periodic weekday effects estimated from 37 different 
data windows $D_{1:84}$ and $D_{t:(t+41)}$ with 
$t \in 1\!:\!36$. All estimated effects are strongly negative on the weekend, but only slightly higher on Mondays than on other working days. It seems that a majority of the cases not reported on weekends are spread randomly over the remaining five days. There is also an indication of small trends in the weekday effects over time, most clearly for Tuesday and Saturday. It is not clear how to incorporate time-varying weekday effects in $v$ or $w$ and how to modify the weekday effects in weeks with additional public holidays.

\section{The EM algorithm}
\label{app-em}
Under the assumption that the times between infection and detection are
independent for different individuals it follows that
$$E[D_t \mid \mb I] = \sum_{s=t-K_m}^{t-1} I_s m_{s,t}.$$
Hence a moment estimator $\hat{\mb I}$ of $\mb I$ should satisfy
\begin{equation}
  \label{eq:em}
  \sum_{s=t-K_m}^{t-1} \hat{I}_s m_{s,t} = D_t \quad (t=1, \ldots, T)
\end{equation}
which is a convolution equation in the time-invariant case
$m_{s,t}=m_{t-s}$. However, this equation has, in general, many solutions,
since the number of unknowns is $T+K_m-1$ and the number of observations
is $T$. 

The linear space of all solutions has dimension $K_m-1$ and one can give an explicit expression
for the general solution if $T>K_m$ and $m_{s,s+K_m}>0$ for all $s$. This becomes clear if we write \eqref{eq:em} as
$$\hat{I}_{t-K_m} m_{t-K_m,t}=
  D_t - \sum_{s=t-K_m+1}^{t-1} \hat{I}_s m_{s,t}, \quad t=1, \ldots, T.$$
Hence for any choice of
$(\hat{I}_s; T-K_m+1 \leq s \leq T-1)$ there is a unique  solution of \eqref{eq:em} that can be computed recursively for for $s=T-K_m, T-1-K_m, \ldots 1-K_m$.

The EM algorithm, also called the Richardson-Lucy deconvolution, is an
iterative algorithm where the $k$-th iteration is defined as
$$\hat{I}^{(k)}_s = \hat{I}^{(k-1)}_s \cdot \frac{1}{b_s} \sum_{t=1}^T m_{s,t}
\frac{D_t} {E[D_t \mid \hat{\mb I}^{(k-1)}]} \quad (s=1-K_m, \ldots, T-1)$$
The starting value $\hat{\mb I}^{(0)}$ is arbitrary. It is obvious that
all iterations remain positive if the starting value is positive. Moreover,
if the iterations converge, the limit is a solution of the deconvolution
equation provided $m_{s,s+K_m}>0$ for all $1-K_m \leq s \leq T-K_m$
because the matrix $m$ is lower triangular.

The EM algorithm is in fact an iterative maximizer of
$$\ell(\mb I) = \sum_{t=1}^T(- E[D_t \mid \mb I] +
D_t \log E[D_t \mid \mb I])$$
which is the log likelihood of $\mb I$ if the variables $D_t$ are independent and Poisson-distributed with mean $\mb I$. It is concave and thus every local maximum
is a global maximum and the set of maximizers is convex. The acronym EM
stands for expectation-maximization: The $k$-th iterate $\mb I^{(k)}$
maximizes
$$\ell^{(k)}(\mb I) = \sum_{s=1-K_m}^{T-1}\left( - I_s \sum_{t=1}^T m_{s,t} +
\log(I_s) \cdot \hat{I}_s^{(k-1)} \sum_{t=1}^T m_{s,t} 
\frac{D_t}{E[D_t \mid \hat{\mb I}^{(k-1)}]}\right).$$
This is the expected log likelihood of $\mb I$ for independent
variables $A_{s,t} \sim$ Poisson($I_s m_{s,t}$), conditional
on $D_t = \sum_s A_{s,t}$ and computed under the current value
$\hat{\mb I}^{(k-1)}$. By Jensen's inequality,
$\ell(\mb I)$ increases strictly for each EM iteration. Therefore
$\ell(\hat{\mb I}^{(k)})$ converges to the global maximum. 

\begin{figure}[ht!]
	\begin{center}
          \includegraphics[width=\textwidth]{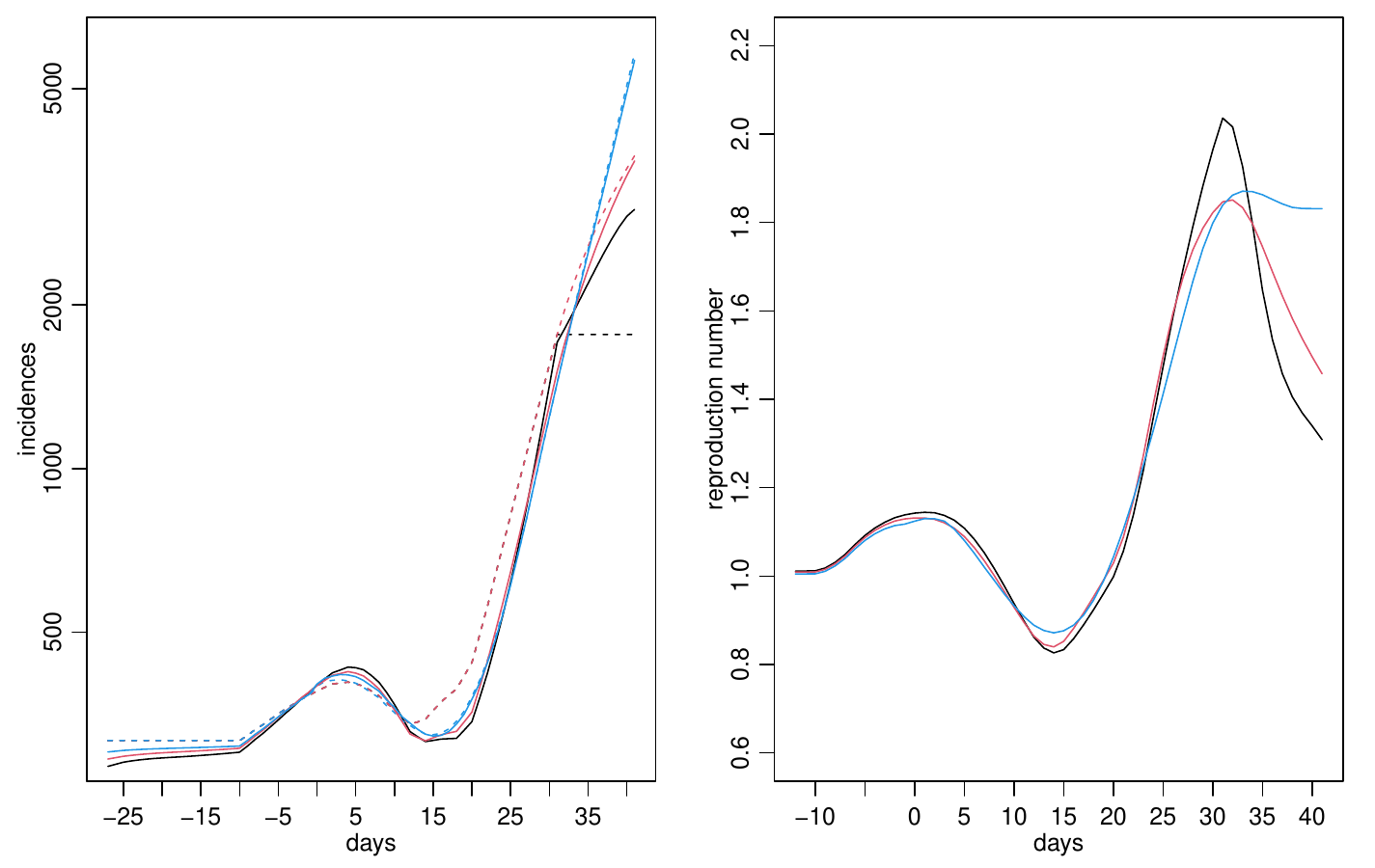}
          \caption{Left: Three starting values (dashed)  and the results of the EM-algorithm (solid). Right: Estimates of reproduction numbers assuming that $I_s$ equals one of the 
          values on the left and $R_{e,s}$ is constant on a sliding window of length 4 days.}
          \label{em-alg}
	\end{center}
      \end{figure}

The EM algorithm is a reasonable procedure to approximate a positive
solution of the convolution equation, but the use of the Poisson likelihood
$\ell(\mb I)$ is not justified. Moreover, convergence of the EM algorithm
can be very slow and the limit can depend quite strongly on the choice of the starting value $\hat{\mb I}^{(0)}$. Fig. \ref{em-alg} shows three plausible starting values and the resulting estimates of $I_s$ and
$R_s$ for $T=42$ when $D_{1:42}$ equals the smoothed detections from Switzerland starting on August 30, 2020. The first starting value shifts
$D_{1:42}$ back by ten days and uses constant extrapolation at both ends. The second starting value uses linear extrapolation on the right, and the third starting value is equal to the median of $\mb I$ from our MCMC sample. In all three cases the EM-algorithm was iterated until the 
chi-squared statistic
$$\sum_{t=1}^{42} (D_t - E[D_t \mid \mb I])^2/E[D_t \mid \mb I]$$
is less than 42. 

For uniqueness one could
add a concave regularization term to $\ell(\mb I)$. \citet{green1990use} proposed
an extension of the EM algorithm when such a regularization term is added,
but our attempts to apply this for our problem were not successful.

\section{Derivation of results mentioned in Section \ref{extension}}
\label{app-ext}

The result for the conditional distribution of $U_s$ given $B_s$
and $\lambda_s$ follows from 
$$p(U_s \mid B_s, \lambda_s) = \frac{p(B_s, U_s \mid \lambda_s)}
{p(B_s \mid \lambda_s)}$$
and the following computations
\begin{eqnarray*}
  p(B_s, U_s \mid \lambda_s) &=& \frac{\eta_s^\gamma b_s^{B_s} u_s^{U_s}}
{\Gamma(\gamma) B_s! U_s!}\int \mu^{\gamma -1 + B_s + U_s}
\exp(- (1+\eta_s)\mu) d\mu\\
&=& \frac{\Gamma(B_s + U_s + \gamma)}{\Gamma(\gamma) B_s! U_s!}
    \left(\frac{b_s}{1+\eta_s}\right)^{B_s} \left(\frac{u_s}{1 +\eta_s}\right)^{U_s}
  \left(\frac{\eta_s}{1 +\eta_s}\right)^\gamma,\\
p(B_s \mid \lambda_s) &=& \frac{\eta_s^\gamma b_s^{B_s}}
{\Gamma(\gamma) B_s!}\int \mu^{\gamma - 1 + B_s}\exp(- (b_s+\eta_s)\mu) d\mu\\
&=& \frac{\Gamma(B_s + \gamma)}{\Gamma(\gamma) B_s!}
\left(\frac{b_s}{b_s+\eta_s}\right)^{B_s}\left(\frac{\eta_s}{b_s+\eta_s}\right)^\gamma
  \end{eqnarray*}

The formula for
the acceptance probability of a proposal $(\mb A^*, \mb I^*)$ is obtained from
\begin{eqnarray*}
  \log p(\mb I^*_{init}, \mb I^*,\mb A^* \mid \mb L, \mb D)
  &=& \log p(\mb I^*_{init}) - \log p(\mb D \mid \mb L)\\                                               
  &+& \sum_{s=1-K_m}^{T-1} \left(\log\Gamma(I^*_s+\gamma) - \log \Gamma(\gamma)
      + I^*_s\log \frac{\lambda^*_s}{\gamma + \lambda^*_s} +
      \gamma\log \frac{\gamma}{\gamma + \lambda^*_s} \right)\\
  &+& \sum_{s=1-K_m}^{T-1} \left(\sum_{t=1}^T(A^*_{s,t} \log m_{s,t} -
      \log A_{s,t}^*!) + U^*_s \log u_s - \log U^*_s! \right)
\end{eqnarray*}
and
\begin{eqnarray*}
  \log q(\mb I^*_{init}, \mb I^*,\mb A^* \mid \mb L, \mb D, \mb I)
  &=& \sum_{t=1}^T \left( \log D_t! +
      \sum_{s=t-K_m}^{t-1} (A^*_{s,t} \log \nu^*_{s,t} - \log A^*_{s,t}!)\right)
      + \log p(\mb I^*_{init})\\
  &+& \sum_{s=1-K_m}^{T-1}\left( \log\Gamma(U^*_s + B^*_s + \gamma) -
      \log\Gamma(B^*_s + \gamma) - \log U^*_s! + \right.\\
  && \hspace{1.5cm} \left. U^*_s\log\frac{u_s \lambda^*_s}
     {\gamma + \lambda^*_s} + (B^*_s + \gamma)
      \log\frac{b_s \lambda^*_s + \gamma}{\gamma + \lambda^*_s}\right).
\end{eqnarray*}
From this and \eqref{eq:acc-ratio-1}, we obtain
\begin{eqnarray*}
  \lefteqn{\log p(\mb I^*,\mb A^*, \mb I^*_{init} \mid \mb L, \mb D) -
  \log q(\mb I^*,\mb A^*, \mb I^*_{init} \mid \mb L, \mb D, \mb I)}\\
  &=& \sum_{t=1}^T (D_t \log \pi^*_t - \log D_t!) - 
      \sum_{s=1-K_m}^{T-1}(B^*_s\log \psi^*_s + \log B^*_s!)\\
  &+& \sum_{s=1-K_m}^{T-1}\left(\log \Gamma(B^*_s + \gamma) - \log \Gamma(\gamma)
      - \log B^*_s! + B^*_s \log\frac{\lambda^*_s}{b_s \lambda^*_s + \gamma}
      + \right.\\
   &&  \hspace{1.5cm} \left. \gamma \log \frac{\gamma}{b_s \lambda^*_s + \gamma}
      + U^*_s (\log u_s + \log\frac{\lambda^*_s}{u_s \lambda^*_s}) \right).
\end{eqnarray*}
Using \eqref{eq:acc-ratio-2},
we arrive at
 \begin{eqnarray*}
  \lefteqn{\log p(\mb I^*,\mb A^*, \mb I^*_{init}\mid \mb L, \mb D) -
  \log q(\mb I^*,\mb A^*, \mb I^*_{init} \mid \mb L, \mb D, \mb I)}\\
   &=& \sum_{t=1}^T \log p_{Pois}(D_t,\pi^*_t) + \sum_{s=1-K_m}^{T-1}
       (p_{neg-binom}(B^*_s; \gamma, \frac{\gamma}{b_s \lambda^*_s + \gamma})
      - p_{Pois}(B^*_s, b_s\psi^*_s)).
\end{eqnarray*} 

Under the model \eqref{eq:model-I-ext}, $\log p(\mb L, \mb I)$ is 
up to additive terms that do not contain $\mb L$ equal to 
$$-\frac{L_{1-K_m}^2}{2 \sigma^2} - \sum_{s=1-K_m}^{T-2} \frac{(L_{t+1}-L_t)^2}
{2 \tau^2} + \sum_{s=1-K_m}^{T-1} (I_s L_s - (I_s + \gamma) \log(\lambda_s + \gamma)).$$
For proposing a new value $\mb L^*$ given $\mb I$ and the current value $\mb L$, we use the second-order Taylor approximation of
$\log(\lambda^*_s + \gamma)$ at $\lambda_s$:
\begin{eqnarray*}
    \log(\lambda^*_s + \gamma) &\approx& \log(\lambda_s + \gamma)
    + c_{1,s} (L^*_s - L_s) + c_{2,s} (L^*_s - L_s)^2/2.\\
    && c_{1,s}=\lambda_s/(\lambda_s + \gamma), \quad
    c_{2,s} = \gamma \lambda_s/(\lambda_s + \gamma)^2.
\end{eqnarray*}
This then leads to a Gaussian proposal distribution of the form
\eqref{eq:logq-L} where
$$Q(\mb L) = Q_L + \diag(\mb I + \gamma)\mb c_2$$
where $Q_L$ is the precision matrix of the prior for $\mb L$
and 
$$b(\mb L) = \mb I - (\mb I + \gamma)(\mb c_1 - \mb c_2 \mb L).$$

\section{Details on the code of \citet{huisman2022estimation}}
\label{app_sub_implement}

In their code,
\citet{huisman2022estimation} drop the $10$ most recent days and go
back only $5$ days prior to the first observation after doing the
deconvolution for estimating $I$. Further, another $5$ days are
dropped at the beginning of the observation period for estimating $R$. This occurs at the following locations:
\begin{itemize}
\item The amount of truncation of $I_s$ for the most recent days is determined by the variable \texttt{initial\_delta} defined at the
following location: \url{https://github.com/covid-19-Re/shiny-dailyRe/blob/9c428e0a91dbf047251b0638bbc5a682572ad566/app/otherScripts/2_utils_getInfectionIncidence.R\#L545}. 
\item The code at the following location implies that $I$ is only
estimated for $-4\leq t$:
\url{https://github.com/covid-19-Re/shiny-dailyRe/blob/9c428e0a91dbf047251b0638bbc5a682572ad566/app/otherScripts/2_utils_getInfectionIncidence.R\#L620}
(using their default value \texttt{days\_further\_in\_the\_past}=30).
\item At the following code location, another $5$ days are dropped at
the beginning of the observation window:
\url{https://github.com/covid-19-Re/shiny-dailyRe/blob/9c428e0a91dbf047251b0638bbc5a682572ad566/app/otherScripts/ReCountry.R\#L397}.
\end{itemize}

These truncations avoid the uncertainty due to the starting value for the EM-algorithm illustrated in Section \ref{app-em}, but typically the
estimates at the most recent days are of most interest.

\clearpage
\section{Additional plots for Subsection \ref{sub-long}}\label{app_sub-long}
\begin{figure}[ht!]
	\begin{center}
          \includegraphics[width=\textwidth]{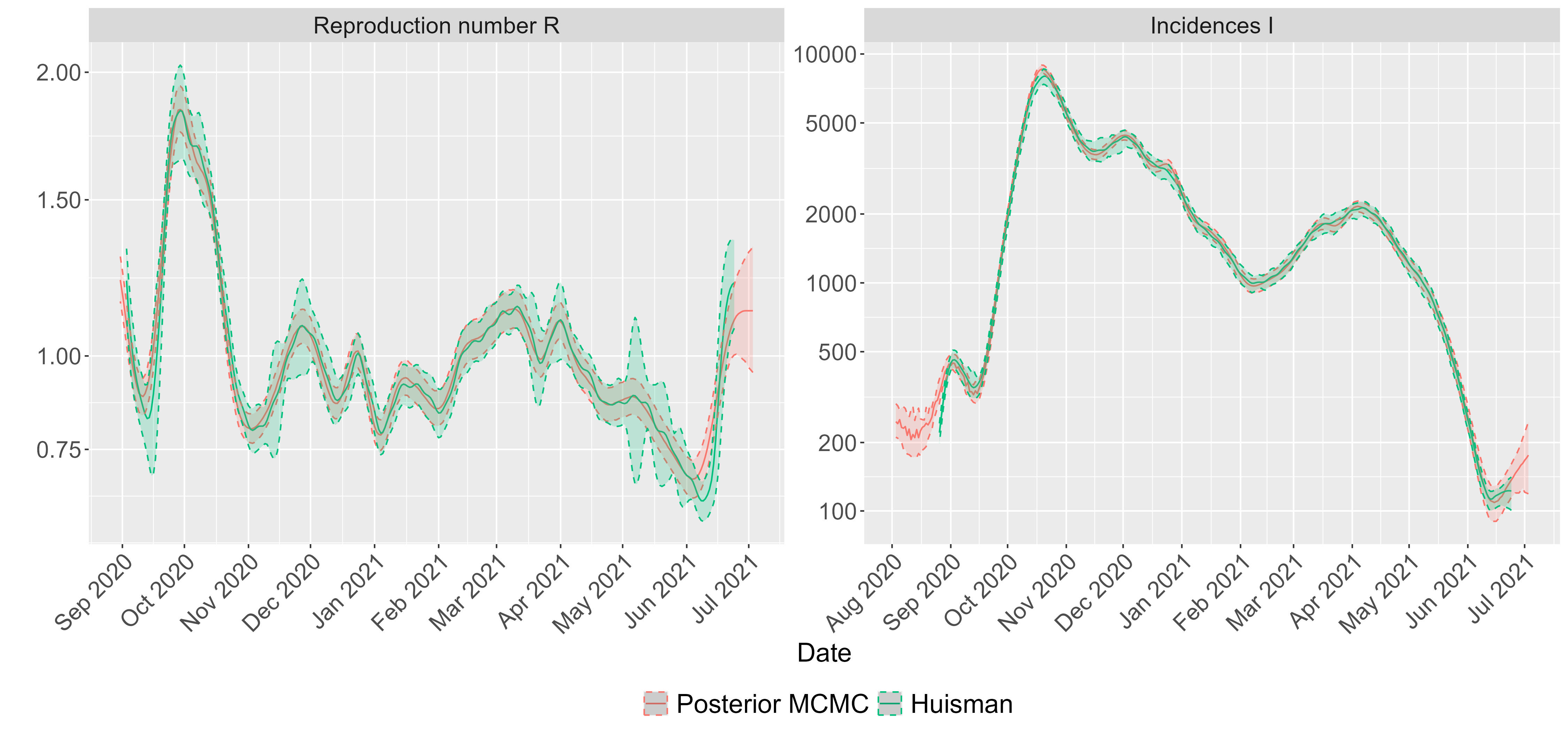}
          \caption{Results when preprocessing the detection data
            using local polynomial regression as smoothing approach. }
          \label{post_loess}
	\end{center}
      \end{figure}

\clearpage
\section{Additional plots for Subsection \ref{simulation}}\label{app_simulation}

\begin{figure}[h!]
  \begin{center}
    \includegraphics[width=0.9\textwidth]{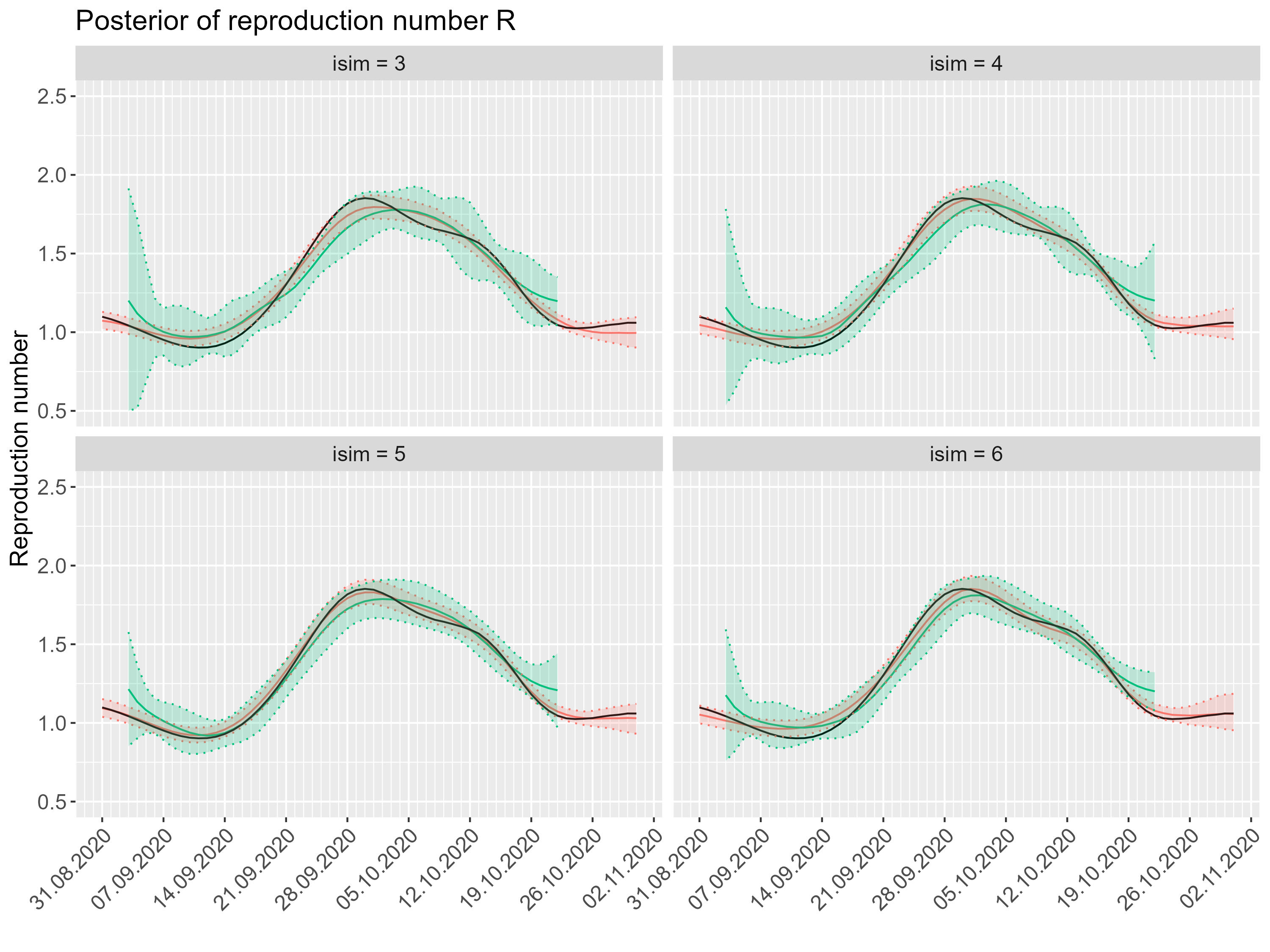}
    \includegraphics[width=0.9\textwidth]{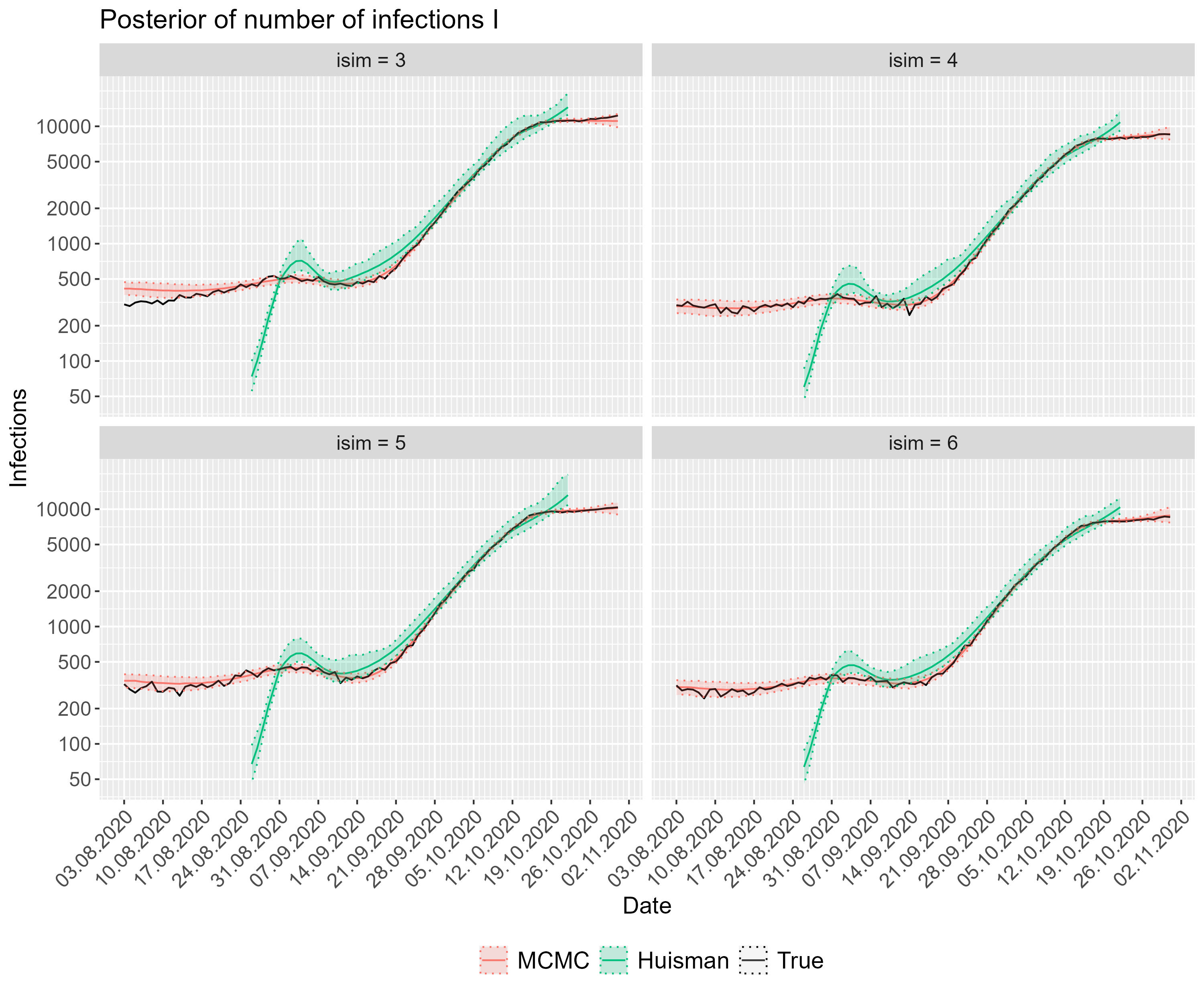}
    \caption{Same as Fig. \ref{post_sim_time} with four additional simulated datasets.} 
		\label{post_sim_time_app}
   \end{center}
\end{figure}

\clearpage 

\begin{figure}[h]
  \begin{center}
      \includegraphics[width=0.9\textwidth]{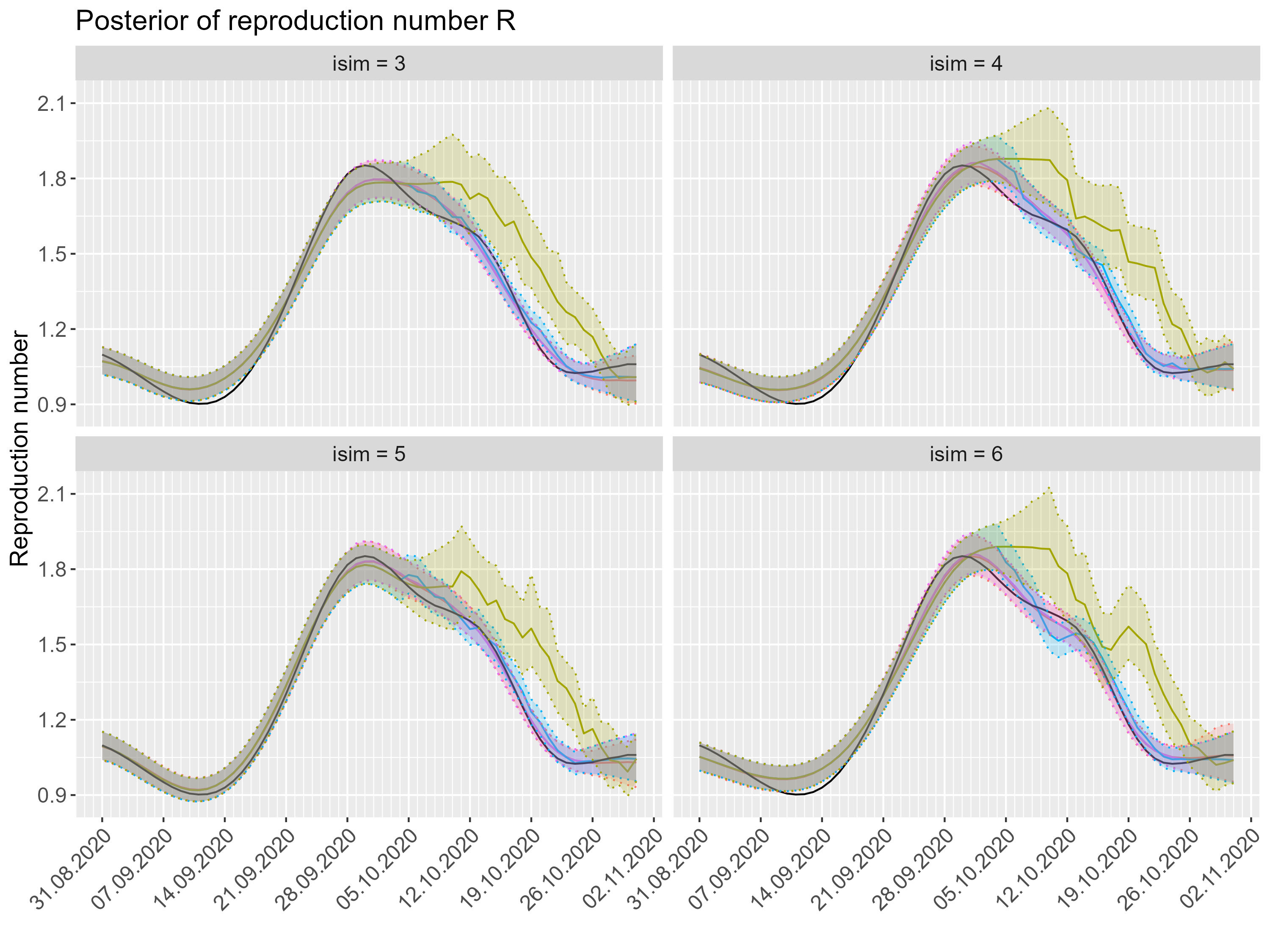}
    \includegraphics[width=0.9\textwidth]{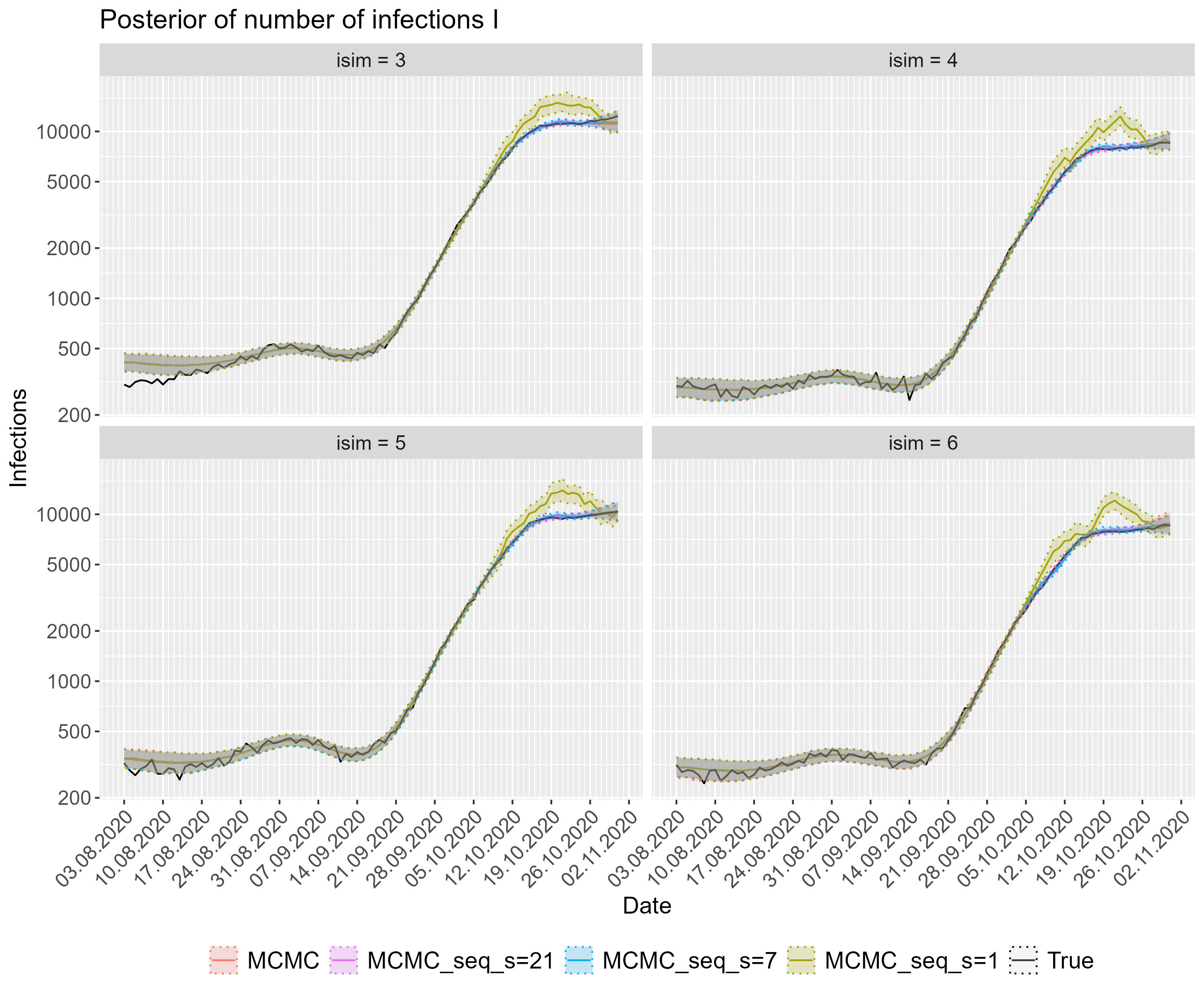}
    \caption{Same as Fig. \ref{post_seq} with four additional simulated datasets.}
 		\label{post_seq_app}
   \end{center}
 \end{figure}

\clearpage
\section{Additional plots for Subsection \ref{sub-short}}
\label{app_sub_short}

\begin{figure}[ht!]
  \begin{center}
    \includegraphics[width=0.9\textwidth]{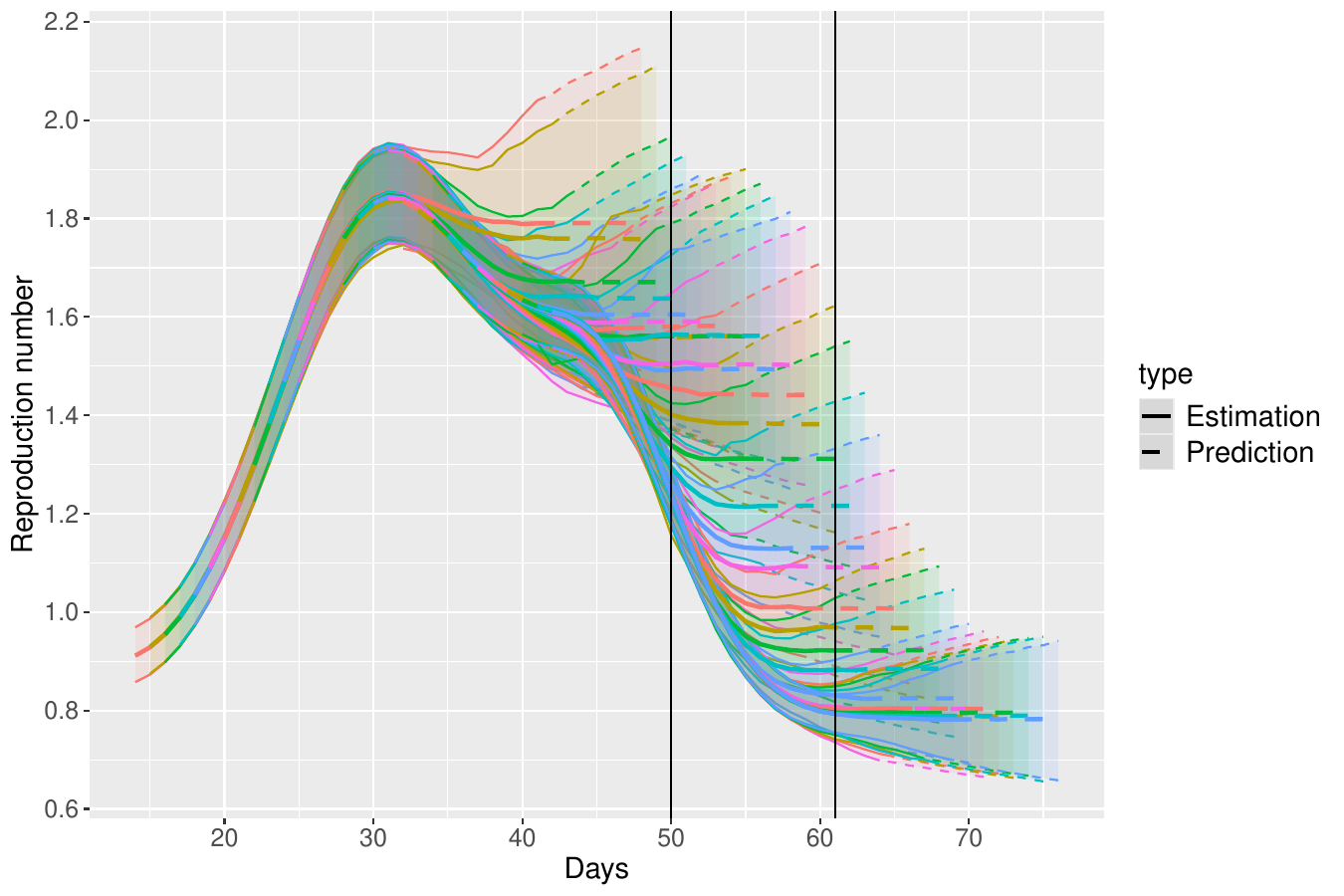}
    \caption{Same as Fig. \ref{r-number-15-77}, but with smooth
      observations computed from raw data $D_{1:84}$.}
    		\label{r-number-15-77-a}
   \end{center}
 \end{figure}
 
\begin{figure}[ht!]
  \begin{center}
    \includegraphics[width=0.9\textwidth]{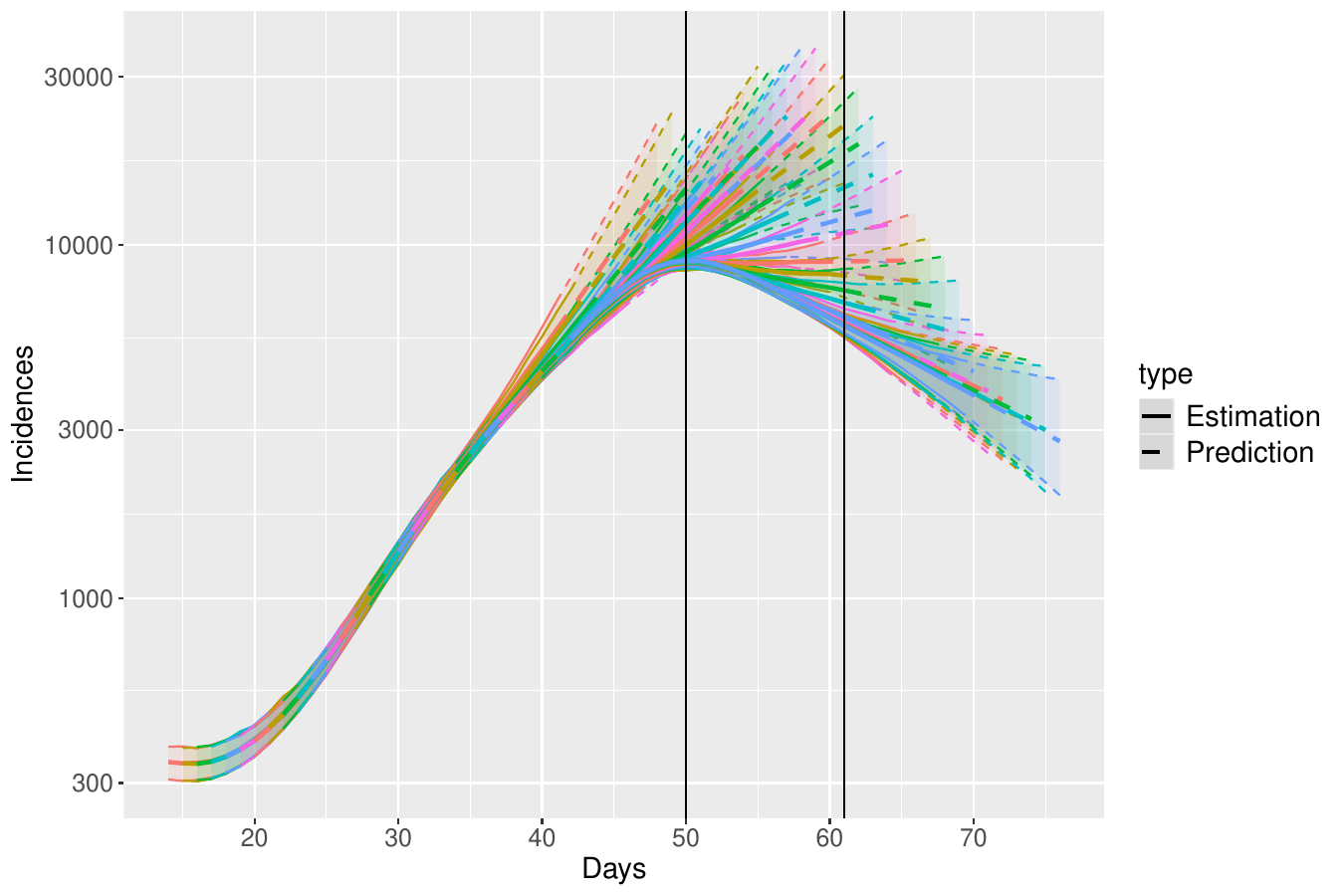}
    \caption{Same as Fig. \ref{infections-15-77}, but with smooth
      observations computed from raw data $D_{1:84}$.}
    		\label{infections-15-77-a}
   \end{center}
 \end{figure}

\begin{figure}[ht!]
  \begin{center}
    \includegraphics[width=0.48\textwidth]{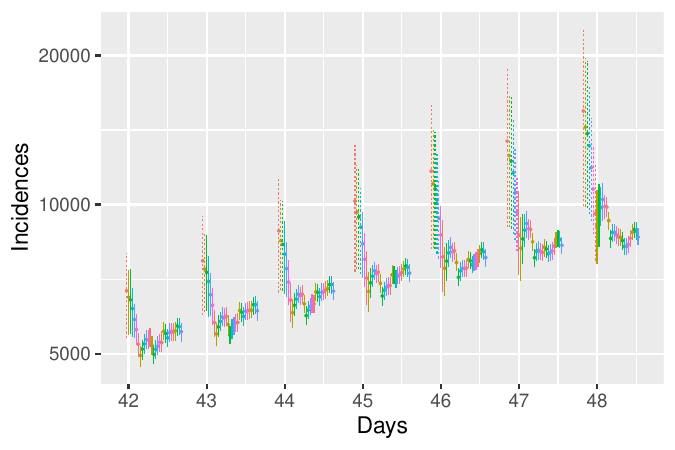}
    \includegraphics[width=0.48\textwidth]{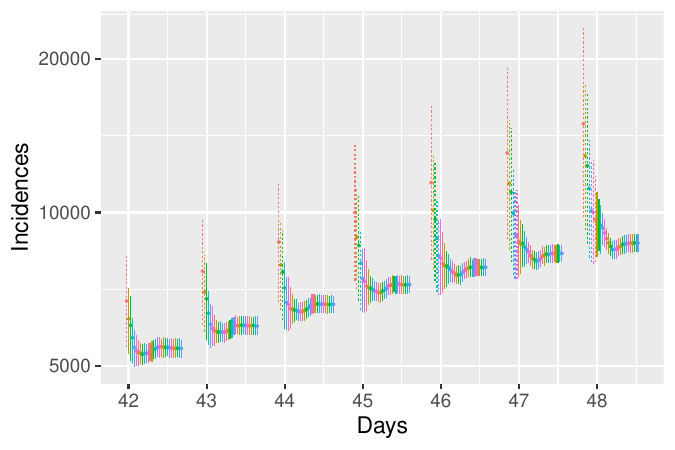}
    \includegraphics[width=0.48\textwidth]{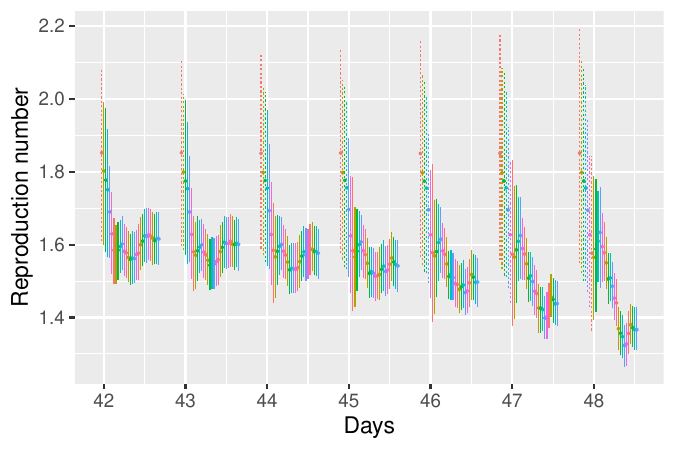}
    \includegraphics[width=0.48\textwidth]{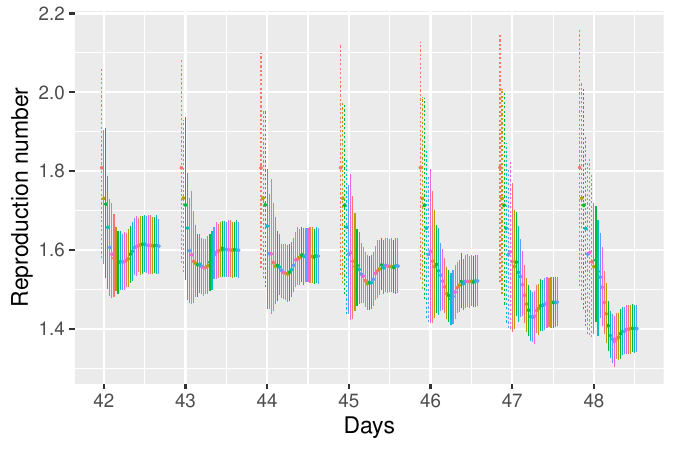}
            \caption{As Fig. \ref{infections-49-55}, but for days $42\!:\!48$.}
    		\label{infections-42-48}
   \end{center}
 \end{figure}

 \begin{figure}[ht!]
  \begin{center}
    \includegraphics[width=0.48\textwidth]{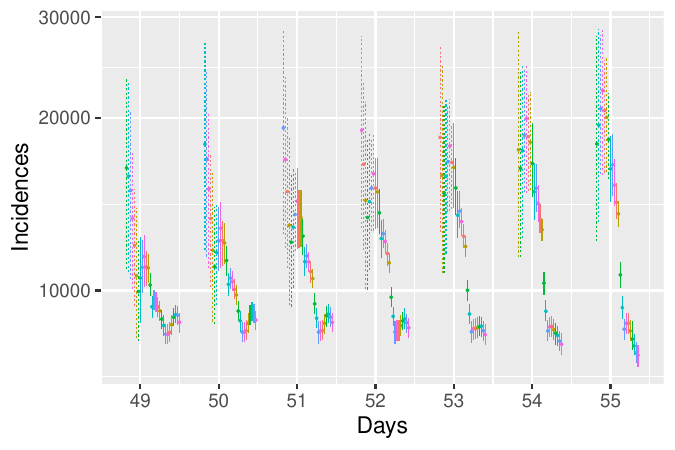}
    \includegraphics[width=0.48\textwidth]{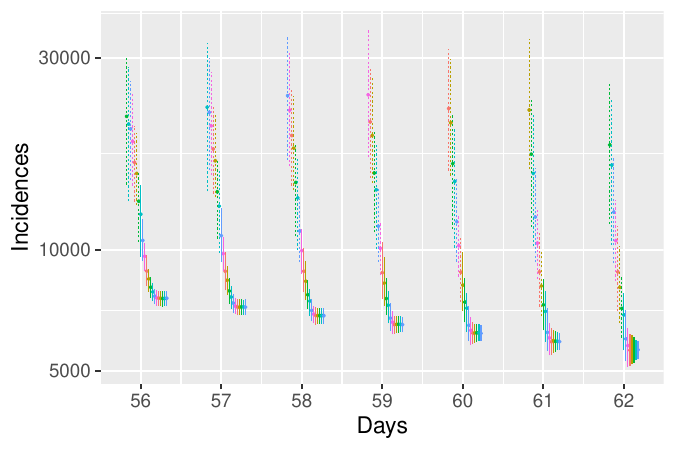}
    \includegraphics[width=0.48\textwidth]{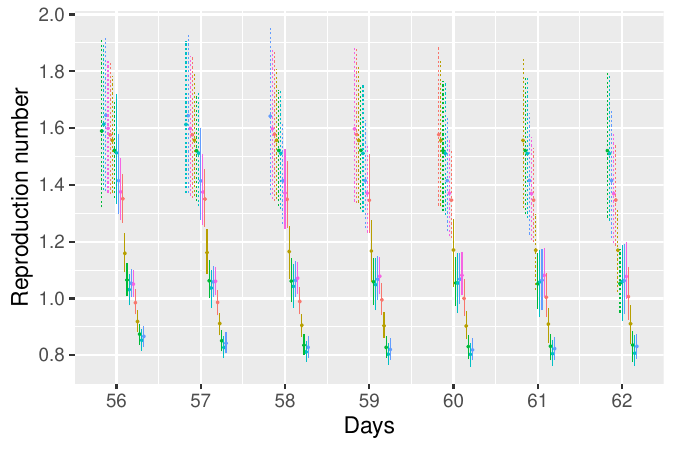}
    \includegraphics[width=0.48\textwidth]{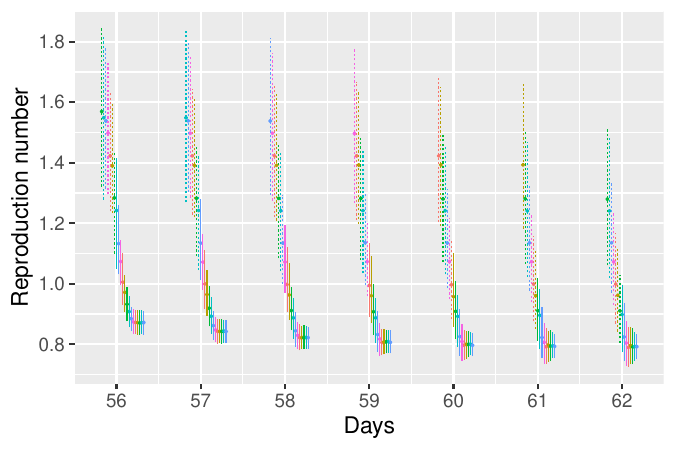}
            \caption{As Fig. \ref{infections-49-55}, but for days $56\!:\!62$.}
    		\label{infections-56-62}
   \end{center}
 \end{figure}

Figs. \ref{infections-42-48} and \ref{infections-56-62} show posterior quantiles of $I_s$ and $R_s$ for $s \in 42\!:\!48$ and $s \in 55\!:\!62$, respectively, slightly shifted horizontally according to the different observation windows. The behavior is similar to what was observed for $s \in 49\!:\!55$,  but the overlap of intervals for different conditioning windows decreases as $s$ increases. Again the overlap is smaller if the smoothed data $\bar D_t$ are computed in the same moving window as used for conditioning. Moreover, two large shifts of the intervals for the reproduction numbers occur in this period when observations on days 61-62 and on days 67-68 become available.

If future data are used for smoothing, observations up to day 63 are needed to confirm that incidences and reproduction numbers continue to decrease in the period $56\!:\!62$ and the reproduction numbers stay below 1. If only data from the same window are used for smoothing, four or five more days are necessary. Thus, even though the epidemic started to slow down well before the interventions occurred, they might have helped to ensure that the reproduction numbers went down below 1 and stayed there.

\end{appendix}